\documentclass[prl,reprint,groupedaddress,showpacs,floatfix]{revtex4-1}
\usepackage[utf8]{inputenc}
\usepackage[pdftex]{graphicx} 
\usepackage{bm} 
\usepackage{amssymb} 
\usepackage{amsmath} 
\usepackage{hyperref}
\usepackage{csquotes}
\usepackage{etoolbox}
\usepackage{float}
\apptocmd{\sloppy}{\hbadness 10000\relax}{}{}
\begin{document}
\title{Exact non-adiabatic part of the Kohn-Sham potential and its fluidic approximation}
\date{\today}
\author{M.\ T.\ Entwistle}
\author{R.\ W.\ Godby}
\affiliation{Department of Physics, University of York, Heslington, York YO10 5DD, United Kingdom\\ and European Theoretical Spectroscopy Facility}

\begin{abstract}   
We present a simple geometrical ``fluidic'' approximation to the non-adiabatic part of the Kohn-Sham potential, $v_{\mathrm{KS}}$, of time-dependent density functional theory. This part of $v_{\mathrm{KS}}$ is often crucial, but most practical functionals utilize an adiabatic approach based on ground-state DFT, limiting their accuracy in many situations. For a variety of model systems, we calculate the exact time-dependent electron density, and find that the fluidic approximation corrects a large part of the error arising from the ``exact adiabatic'' approach, even when the system is evolving far from adiabatically. 
\end{abstract}

\maketitle

Time-dependent Kohn-Sham density functional theory \cite{TDDFT, TDDFT2, KS} (TDDFT) is in principle an exact and efficient theory of the dynamics of systems of interacting electrons. In practical applications, while performing well in some cases, its validity is often restricted by the limitations of available approximate functionals for electron exchange and correlation (xc). Typically, an adiabatic approximation to the xc potential is used, in which the instantaneous electron density is implicitly assumed to be in its ground state, thereby neglecting all \enquote{memory effects}. While these ground-state approximations have steadily improved \cite{KS, PZ_LDA, PW_LDA, GGA, PW91, PW_GGA, GGA2, Hybrid_GGA, Meta_GGA, Meta_GGA2, Meta_GGA3, SCAN}, by definition they cannot approach the exact TDDFT potential: it is necessary to address the non-adiabatic contributions in order for TDDFT to be capable of predictive accuracy in relation to a multitude of applications to diverse fields such as the determination of electronic excitation energies including those of a charge-transfer nature \cite{Adiabatic_fail10}, electron dynamics \cite{Elliot_2012} including non-perturbative charge transfer dynamics \cite{Adiabatic_fail9}, time-resolved spectroscopy \cite{TR_spectroscopy} and electron scattering \cite{Electron_scattering}. 

In this paper, in order to clearly distinguish  between adiabatic and non-adiabatic contributions, we consider the purest application of the concept of the adiabatic functional to the \textit{complete} Kohn-Sham (KS) potential, $v_{\mathrm{KS}}$: at each instant, the DFT KS potential whose \textit{ground-state} density is equal to the exact time-dependent density. The remainder of the exact $v_{\mathrm{KS}}$ constitutes the unambiguously non-adiabatic part, to which we also propose an approximation.

We work in the Runge-Gross formalism \cite{TDDFT} of TDDFT, in which the exact xc potential, $v_{\mathrm{xc}}$, at time $t$ \footnote{The real system of interacting electrons is mapped onto an auxiliary system of noninteracting electrons moving in the effective potential $v_{\mathrm{KS}}$.} depends on the density at all points in space and all non-future times. It has been argued \cite{HPT, zeroforce, fxc_sumrule, VK} that the exact non-adiabatic functional often requires strong nonlocal temporal and spatial dependence on the density. A number of properties of the exact functional, such as the harmonic potential theorem (HPT) \cite{HPT} and zero-force theorem (ZFT) \cite{zeroforce}, have been used to identify limitations of previous approximate TDDFT functionals. Adiabatic functionals trivially satisfy many of these exact conditions through their complete lack of memory-dependence, yet prove inadequate in many applications \cite{Adiabatic_fail, Hessler_2002, Adiabatic_fail2, Adiabatic_fail3, Adiabatic_fail4, Adiabatic_fail5, Adiabatic_fail6, Adiabatic_fail7, Adiabatic_fail8, Maitra_2016, Adiabatic_fail9, Adiabatic_fail10, Adiabatic_fail11, Elliot_2012, TR_spectroscopy, Electron_scattering}. The development of non-adiabatic functionals that continue to satisfy these exact properties is non-trivial. For example, it was shown that modifying the adiabatic local density approximation (ALDA) by introducing time-nonlocality, such as in the Gross-Kohn \cite{GK} (GK) approximation, is inappropriate \cite{zeroforce, HPT}. 

The best-known approximate non-adiabatic functional is that developed by Vignale and Kohn \cite{VK,VK2, VK_fluid} (VK). This was constructed by studying the responses to slowly-varying perturbations of the homogeneous electron gas, and they found a time-dependent xc vector potential as a functional of the local current and charge densities $j$ and $n$, thereby implicitly obtaining a scalar potential which depends nonlocally on the density. While the VK formalism has proved promising \cite{VK_good, VK_good2, VK_good3, VK_good4, VK_good5, VK_good6, VK_good7, VK_good8, VK_good9, VK_good10, VK_good11}, not least through it obeying the HPT and ZFT, its validity is limited \cite{VK_bad, VK_bad2, VK_bad3, VK_bad4, VK_bad5} owing to the constraints under which it was derived.

Our calculations employ the iDEA code \cite{iDEA} which solves the many-electron Schr{\"o}dinger equation exactly for small, one-dimensional prototype systems of spinless electrons \footnote{The use of spinless electrons gives access to richer correlation for a given number of electrons. The electrons interact via the appropriately softened Coulomb repulsion \cite{Thesis_Hodgson} $u(x,x') = (|x-x'|+1)^{-1}$. We use Hartree atomic units: $m_{e}=\hbar=e=4\pi \varepsilon_{0}=1$.} \footnote{ See Supplemental Material at [URL will be inserted by publisher] for the parameters of the model systems, and details of the convergence.}. This gives us access to the exact electron density $n(x,t)$. We then determine the exact $v_{\mathrm{KS}}(x,t)$ through reverse engineering \cite{iDEA_RE}. We also obtain the \textit{exact adiabatic} KS potential \cite{Hessler_2002,AE_Thiele,Maitra_2016} $v_{\mathrm{KS}}^{\mathrm{A}}$ by applying \textit{ground-state} reverse engineering to the instantaneous density at each time \footnote{Our graphs show the various adiabatic and non-adiabatic KS potentials, etc., evaluated on the \textit{exact} time-dependent density, so that any errors in the potentials or densities are entirely attributable to errors in the functionals, not the input to the functionals.}. The \textit{exact non-adiabatic} component $\Delta v_{\mathrm{KS}}$ is then $v_{\mathrm{KS}}-v_{\mathrm{KS}}^{\mathrm{A}}$.

In developing an approximation to $\Delta v_{\mathrm{KS}}$, it is helpful to consider the situation in different inertial frames, related through a Galilean transformation, as noted by Tokatly \textit{et al.} \cite{Lagrangian, Lagrangian2, Adiabatic_fail6, Lagrangian4, Lagrangian5}. While $v_{\mathrm{KS}}^{\mathrm{A}}$ requires zero correction in any inertial frame when the density is fully static in one of these frames, in the more general case the non-adiabatic corrections to $v_{\mathrm{KS}}^{\mathrm{A}}$ may be expected to be at their smallest in the local, instantaneous rest frame of the density, defined by a transformation velocity of the local velocity field  $u(x,t)=j(x,t)/n(x,t)$. In particular, the effects of acceleration ($\dot{u} \neq 0$) and dispersion ($\partial_x u \neq 0$) have least effect in a frame where $u$ itself is zero \footnote{The rate of change of kinetic energy is proportional to $u\dot{u}$ (as in classical mechanics), and so is smallest when $u$ is zero. Also, if the density is moving with velocity $u$ it will more rapidly encounter a region in which a larger non-adiabatic correction is required.}. Conveniently, introducing a vector potential $A=-u(x,t)$ in the original frame of reference is (apart from an unimportant temporal phase factor) equivalent to a Galilean transformation to the local instantaneous rest frame \footnote{The stated $A$ causes the wavefunction in the original frame to become the wavefunction in the instantaneous rest frame multiplied by $\exp(-iu^{2}t/2)$.} \cite{Lagrangian, Lagrangian2}. As described above, the non-adiabatic correction should be minimal in the latter frame, and here we adopt the simple assumption that it is zero. We term this the \textit{fluidic} approximation. The resulting non-adiabatic correction in the original frame is therefore
\begin{equation} \label{fluidic}
    \Delta v_{\mathrm{KS}}(x,t>0) = -\int_{-\infty}^{x} \frac{\partial}{\partial t} u(x',t>0) \ dx',
\end{equation}
where we have gauge-transformed $A$ into a scalar potential. It is evident that the density-dependence of this $\Delta v_{\mathrm{KS}}$ is nonlocal in both space and time \cite{VK}.

\begin{figure}[htbp]      
\centering
\includegraphics[width=1.0\linewidth]{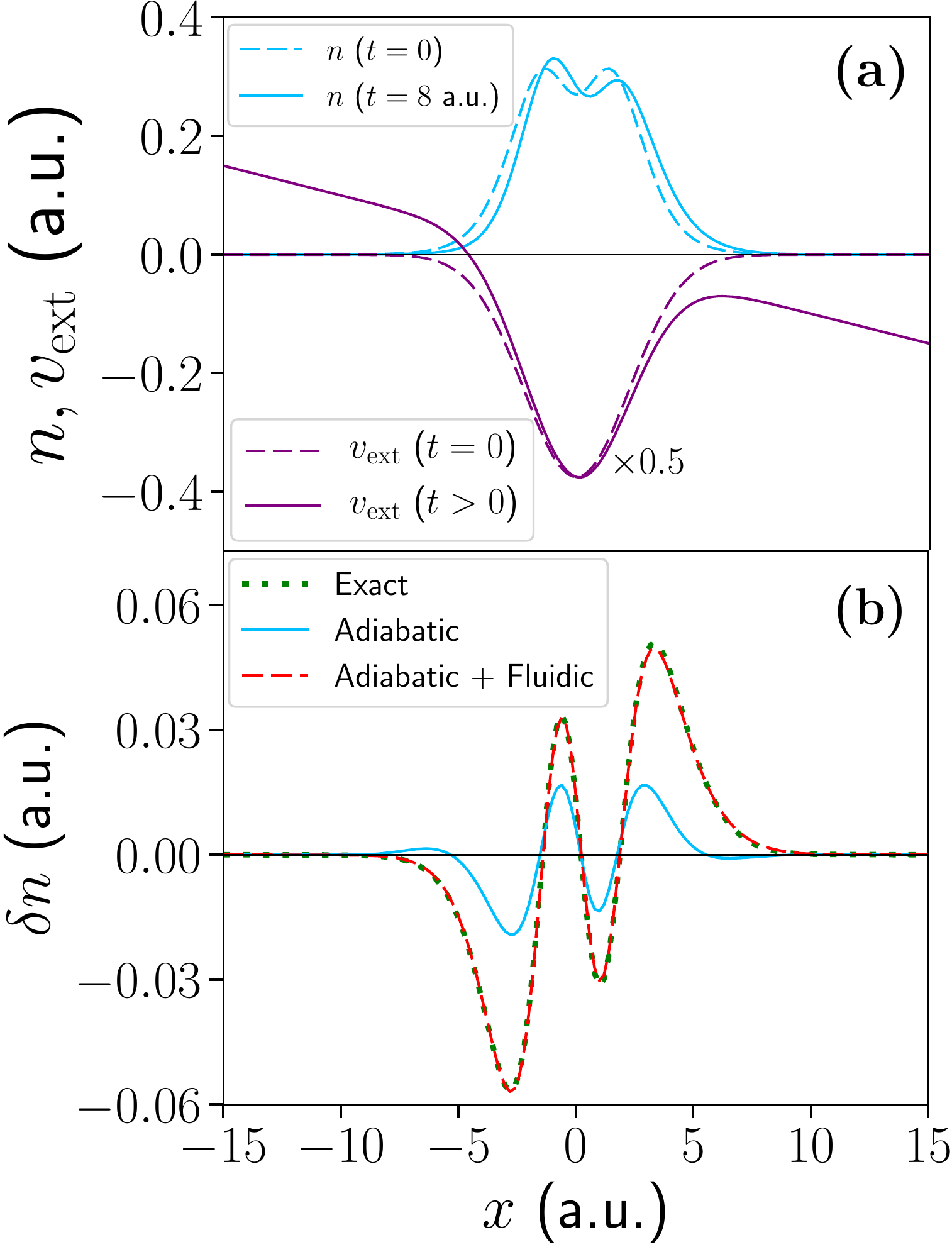}
\caption{System 1: two interacting electrons in a Gaussian potential well, with a uniform electric field applied at $t=0$, driving the electrons to the right and inducing a current. (a) The ground-state external potential (dashed purple) and exact ground-state electron density (dashed blue), along with the perturbed external potential (solid purple) and exact time-dependent electron density at $t=8$ a.u. (solid blue). (b) The change in the exact electron density ($\delta n(x,t) = n(x,t)-n(x,0)$) at $t=8$ a.u. (short-dashed green), along with that  obtained when using the exact $v_{\mathrm{KS}}^{\mathrm{A}}$ (solid blue), and when adding the exact $v_{\mathrm{KS}}^{\mathrm{A}}$ with the fluidic approximation $\Delta A_{\mathrm{KS}} = -u$ (dashed red). The exact adiabatic potential is clearly inadequate, but its error is substantially reduced by the fluidic approximation.}
\label{gaussian_1}
\end{figure}

\textit{System 1} --- As a first test of the fluidic approximation, we consider two interacting electrons in a potential well, which takes the form of an inverted Gaussian function. Initially in the ground state, a uniform electric field, $-\varepsilon x$, is applied at $t=0$, driving the electrons to the right and inducing a current [Fig.~\ref{gaussian_1}(a)]. The sudden application of the perturbation means that we are well outside of the adiabatic limit, and this can be seen by solving the time-dependent KS equations with the exact adiabatic KS potential, $v_{\mathrm{KS}}(t)=v_{\mathrm{KS}}^{\mathrm{A}}(t)$. By plotting the change in the electron density from the ground state, $\delta n$, we find $v_{\mathrm{KS}}^{\mathrm{A}}(t)$ on its own to be wholly inadequate ($\approx 13\%$ error in $n$ \footnote{The integrated absolute error, $\int dx \ |n_{1}(x,t)-n_{2}(x,t)|$, expressed as a percentage of the total number of electrons.} at $t=8$ a.u.), while adding the fluidic approximation substantially reduces this error to less than $1\%$ [Fig.~\ref{gaussian_1}(b)].

\begin{figure}[htbp]      
\centering
\includegraphics[width=1.0\linewidth]{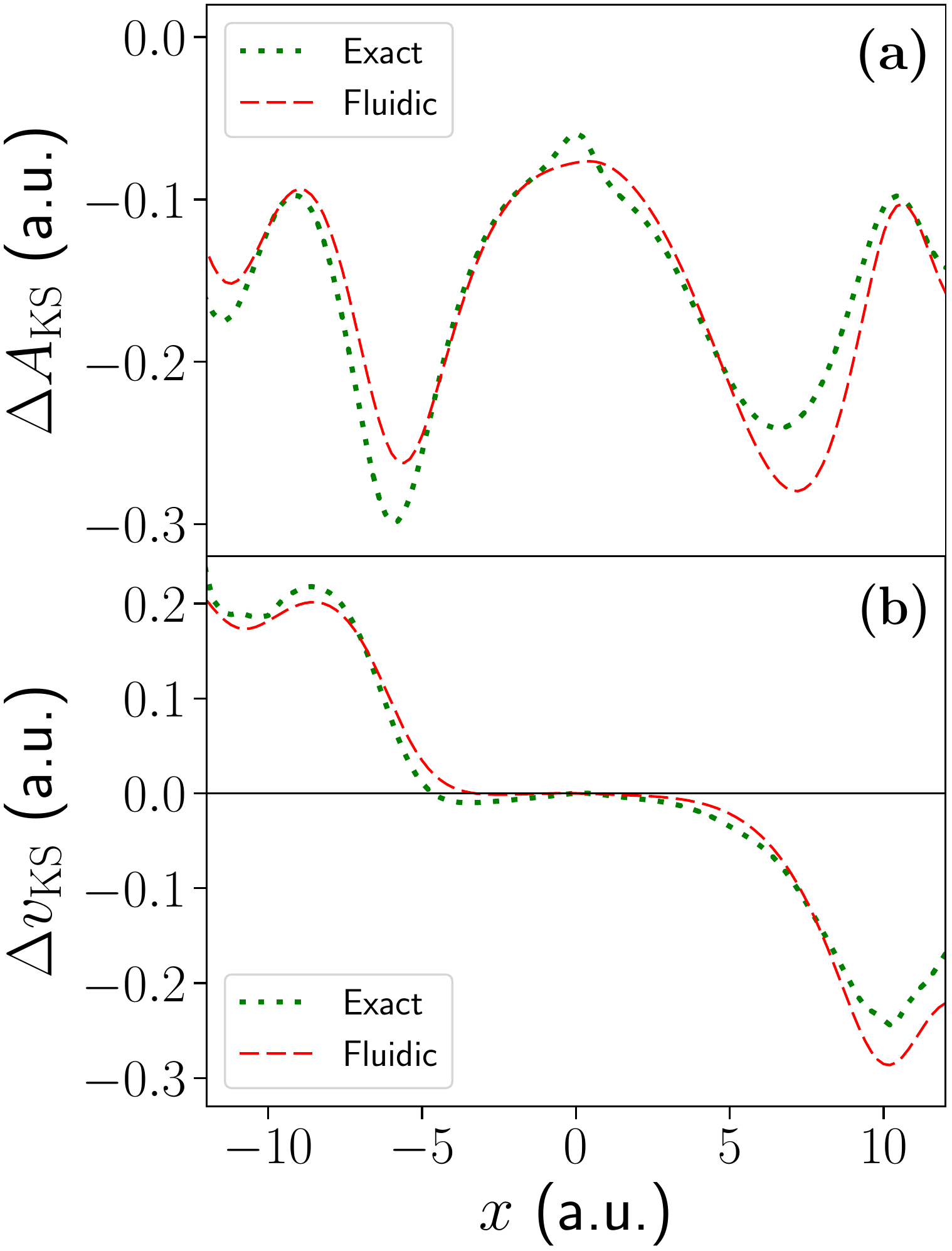}
\caption{The non-adiabatic correction to the KS potentials for System 1. (a) The exact $\Delta A_{\mathrm{KS}}$ (short-dashed green) and that obtained using the fluidic approximation $\Delta A_{\mathrm{KS}} = -u$ (dashed red), at $t=8$ a.u. (b) The corresponding exact (short-dashed green) and fluidic (dashed red) $\Delta v_{\mathrm{KS}}$ in its scalar form. The fluidic approximation performs very well, even though the velocity field is non-uniform in both space and time. (The exact adiabatic approximation, of course, amounts to setting $\Delta A_{\mathrm{KS}}=\Delta v_{\mathrm{KS}}=0$.)}
\label{gaussian_2}
\end{figure}

To understand these results we analyze the non-adiabatic correction to the KS potential in both its scalar and its vector forms. We find very good agreement between the exact $\Delta A_{\mathrm{KS}}$ and that obtained using the fluidic approximation $-u(x,t)$ [Fig.~\ref{gaussian_2}(a)]. The velocity field $u$ (the negative of the fluidic curve in Fig.~\ref{gaussian_2}(a)) quickly becomes strongly non-uniform in both space and time as the electrons explore excited states -- far removed from a universal rest frame. Similarly close agreement between the exact and fluidic $\Delta v_{\mathrm{KS}}$ [Fig.~\ref{gaussian_2}(b)] is evident when the non-adiabatic correction is cast into its scalar form through Eq.~(\ref{fluidic}). 

\textit{Systems 2A, 2B, 2C} --- We now consider a set of systems of interacting electrons in atomiclike external potentials which decay much more slowly at large $x$, $v_{\mathrm{ext}}=-a/(|x|+a)$ with $a=20$, thereby increasing correlation. At time $t=0$, a static sinusoidal perturbation of the form $\varepsilon \cos(0.75x)$ is applied, where  $\varepsilon$ is 0.02 for System 2A (two electrons), 0.02 for System 2B (three electrons) and 0.1 for System 2C (three electrons). 

In System 2A the sudden perturbation at $t=0$ acts to push the two electrons apart. This results in a velocity field that is varying in both space and time, as in System 1; in this case even the sign of $u$ is not the same for all $x$, which takes us even further away from a universal rest frame. Correspondingly, we find the exact adiabatic potential to be insufficient ($\approx 5\%$ error in $n$ at $t=5$ a.u.), while adding the fluidic approximation reduces this error to $\approx 1\%$. System 2B contains three interacting electrons in the same $v_{\mathrm{ext}}$ as System 2A. The additional electron results in a ground-state density that is much less spatially uniform. We run the simulation for 5 a.u. of time and find similar results: $v_{\mathrm{KS}}^{\mathrm{A}}$ produces an error in $n$ of $\approx 5\%$, and the fluidic approximation reduces this to $\approx 1\%$.

\begin{figure}[H]      
\centering
\includegraphics[width=1.0\linewidth]{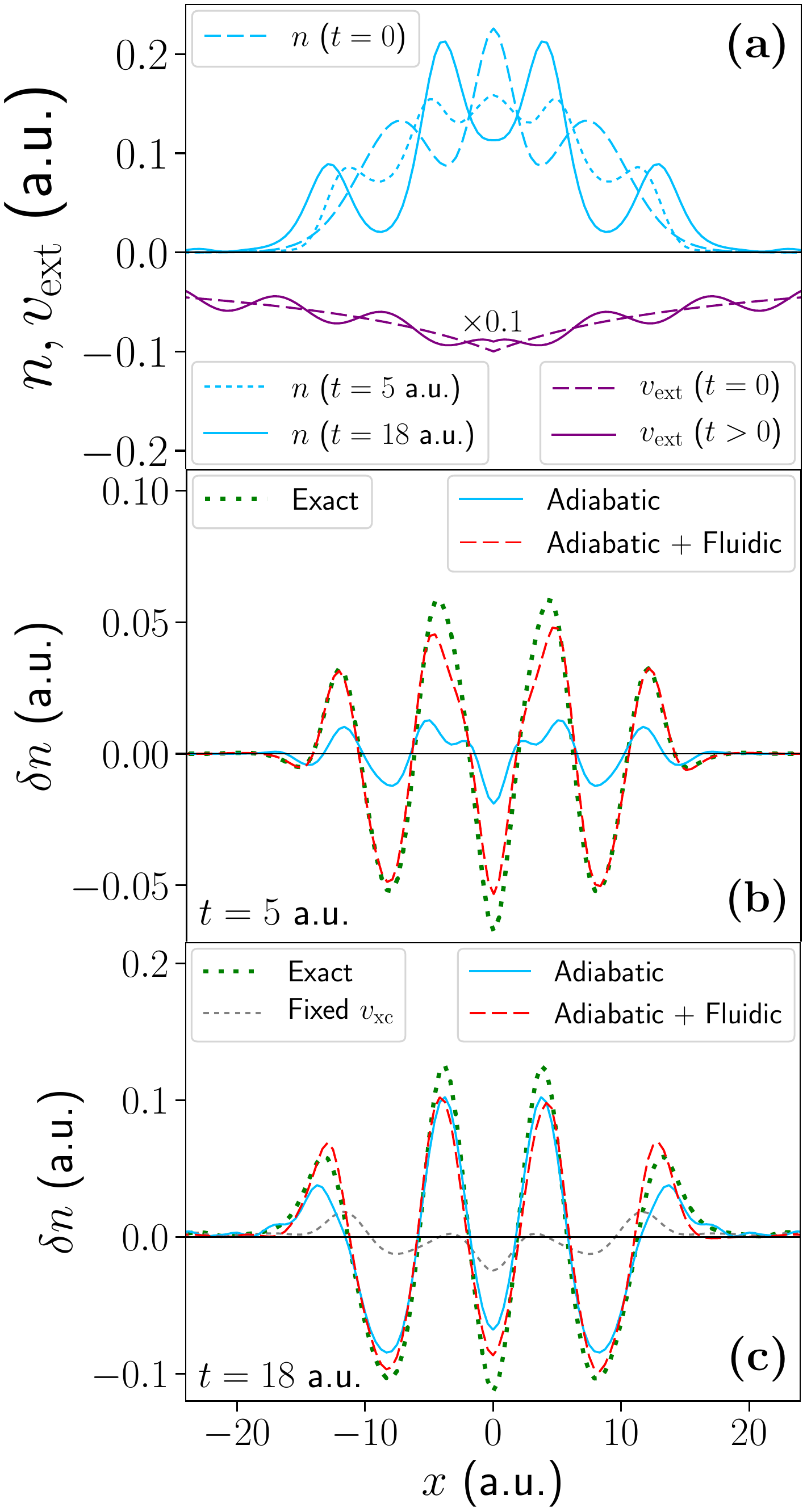}
\caption{System 2C: three interacting electrons in an atomiclike potential, with a static sinusoidal perturbation applied at $t=0$, pushing the electrons apart. (a) The ground-state external potential (dashed purple) and exact ground-state electron density (dashed blue), along with the perturbed external potential (solid purple) and exact time-dependent electron density at $t=5$ a.u. (short-dashed blue) and $t=18$ a.u. (solid blue). (b) The change in the exact electron density at $t=5$ a.u. (short-dashed green), along with that obtained when using the exact $v_{\mathrm{KS}}^{\mathrm{A}}$ (solid blue), and when adding the exact $v_{\mathrm{KS}}^{\mathrm{A}}$ with the fluidic approximation (dashed red). Even though the density is strongly disrupted, the fluidic approximation remains successful. (c) The same as (b) but at $t=18$ a.u. where the dynamic xc contribution is very significant, evident by the completely inadequate result obtained with the fixed $v_{\mathrm{xc}}$ (short-dashed gray) (see main text). Here, the exact $v_{\mathrm{KS}}^{\mathrm{A}}$ is better, but adding the fluidic approximation improves it further.}
\label{atom_1}
\end{figure}

As mentioned above, the fluidic approximation assumes that a system remains close to its ground state in the local instantaneous rest frame. In order to stretch this approximation severely, in System 2C the perturbing potential is much stronger, resulting in a much larger response of the density [Fig.~\ref{atom_1}(a)]. The fluidic approximation still succeeds in reducing the error in the density, from $\approx 25\%$ where only the exact adiabatic potential is used, to $\approx 6\%$, at $t=5$ a.u. [Fig.~\ref{atom_1}(b)]. At later times, the dynamic (time-dependent) xc effects become very significant. To confirm this, we replace the xc component of the exact time-dependent $v_{\mathrm{KS}}$ with the fixed ground-state $v_{\mathrm{xc}}$, thereby suppressing the dynamic part, and find this potential to be wholly inadequate ($\approx 62\%$ error in $n$ at $t=18$ a.u.). Here, the exact adiabatic KS potential is better ($\approx 17\%$ error), while adding the fluidic approximation improves it further ($\approx 15\%$ error) [Fig.~\ref{atom_1}(c)].

\textit{Exact conditions} --- A number of properties of the exact xc functional are known, and these are often used to identify the limitations of approximate functionals. We now explore whether the fluidic approximation satisfies these exact conditions.

We begin with the \textit{one-electron limit}, where the exact xc functional, when applied to a one-electron system, reduces to the negative of the Hartree potential $v_{\mathrm{H}}$, thereby canceling the spurious self-interaction. This means that $v_{\mathrm{KS}}$ is described exactly by a known functional \cite{Hessler_2002, Elliot_2012, Maitra_2016}, which has been termed \cite{iDEA_SOA} the single orbital approximation -- itself capable of capturing features such as steps in the KS potential \cite{Elliot_2012, Thesis_Hodgson} -- whose non-adiabatic part is 
\begin{equation} \label{SOA}
    \Delta v_{\mathrm{KS}}(x,t) = - \int_{-\infty}^{x} \frac{\partial}{\partial t} u(x',t) \ dx' -\frac{1}{2} u^{2}(x,t).  
\end{equation}
We note that the first term is the fluidic approximation [Eq.~(\ref{fluidic})]. We have studied systems of one electron in the external potentials from Systems 1, 2A and 2C, and confirm that the full Eq.~(\ref{SOA}) yields the exact $v_{\mathrm{KS}}$; here, the effect on the density of including the $-u^{2}/2$ term ranges from $<$0.1\% (potential 2A) to 14\% (potential 2C), so that the fluidic approximation alone is already satisfactory. Indeed, in our two- and three-electron systems, the effect of adding the additional term to the fluidic approximation is small and typically slightly deleterious. 

The \textit{zero-force theorem} \cite{zeroforce} follows from Newton's third law and requires the net force exerted on the system by $v_{\mathrm{H}}$ and $v_{\mathrm{xc}}$ to vanish. At the level of the KS potential, $\int n(x,t) \partial_{x} \Delta v_{\mathrm{KS}}(x,t) \ dx = \int n(x,t) \partial_{x} v_{\mathrm{ext}}(x,t) \ dx$, since the exact $v_{\mathrm{KS}}^{\mathrm{A}}$ satisfies the theorem in its own right. In the fluidic approximation for System 1 \footnote{Systems 2A, 2B and 2C satisfy the theorem owing to their symmetry, so do not form a useful test.}, the left and right hand sides of this equation are within 11 \% of one another, so that the theorem appears to be approximately obeyed.

The \textit{harmonic potential theorem} \cite{HPT} shows that in a system of interacting electrons in a harmonic potential, subject to a uniform electric field at $t=0$, the density rigidly moves in the manner of the underlying classical harmonic oscillator. We have shown that the fluidic approximation adds exactly the non-adiabatic correction required \footnote{Apart from an unimportant time-dependent constant.} by the HPT. We have also confirmed this numerically for two interacting electrons in a harmonic potential.

A constraint that can be challenging for non-adiabatic functionals is the \textit{memory condition} \cite{memory_condition}, which notes that $v_{\mathrm{xc}}(t)$ and hence $v_{\mathrm{KS}}(t)$ must be independent of which previous instant in the evolution of the system is to be used to designate the ``initial state''. This is violated by the VK functional \cite{Maitra_2016}.  Eq.~(\ref{fluidic}) demonstrates that the fluidic approximation satisfies this memory condition by virtue of its dependence only on the instantaneous rate of change of $u$, and not its full history.

\begin{figure}[htbp]      
\centering
\includegraphics[width=1.0\linewidth]{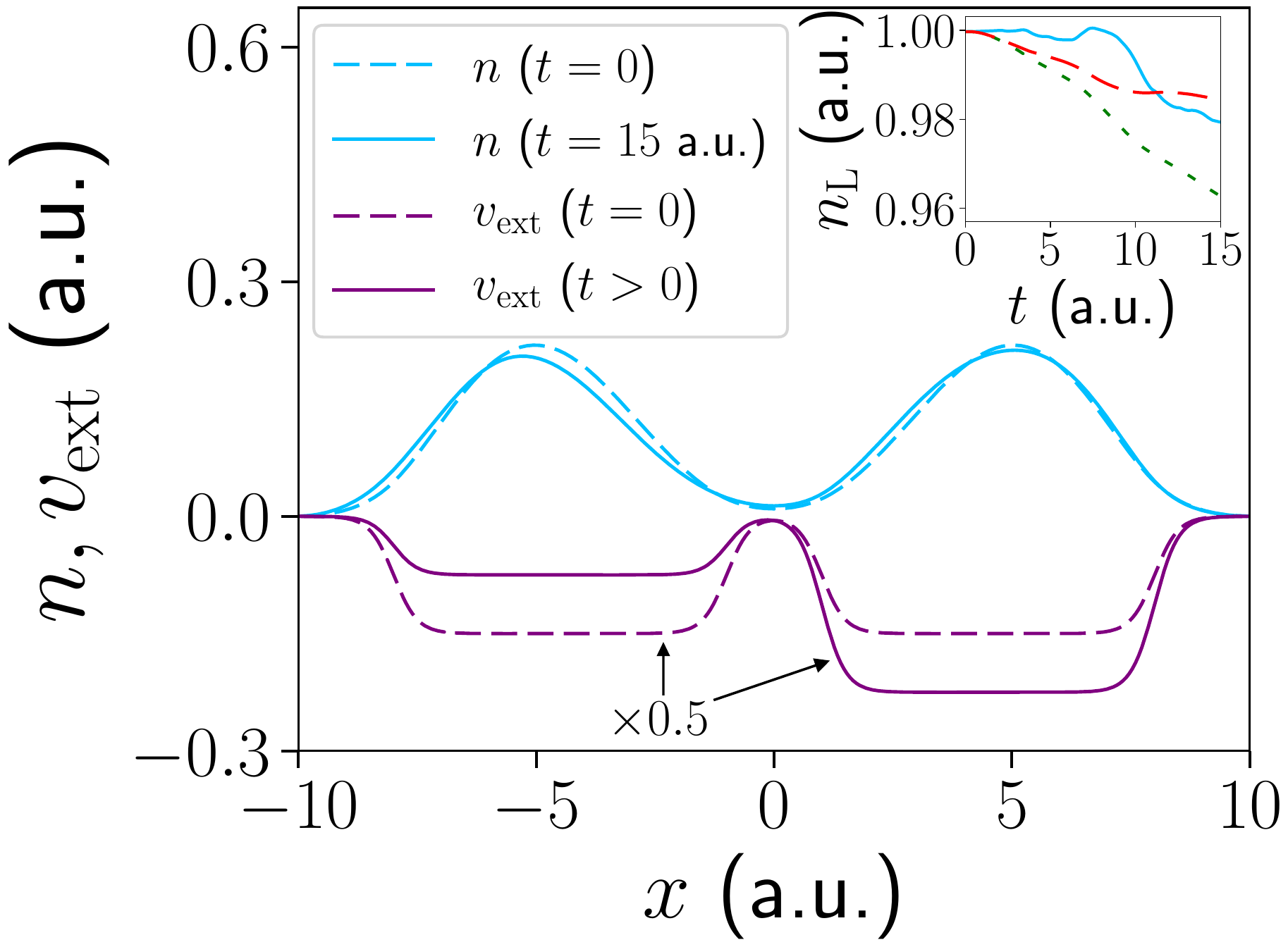}
\caption{System 3: two interacting electrons in a tunneling system. Inset: The exact total electron number on the left-hand side ($x<0$) (short-dashed green); also the exact adiabatic (solid blue) and fluidic approximation (dashed red).}
\label{tunneling_1}
\end{figure}

\textit{System 3} --- As a challenging test of the fluidic approximation, we finally consider two interacting electrons in a tunneling system. Initially $v_{\mathrm{ext}}$ is a symmetric double-well potential, with one electron localized in each well. At $t=0$, the left-hand well is raised and the right-hand well lowered, initiating tunneling through the barrier [Fig.~\ref{tunneling_1}]. A tunneling electron has an imaginary momentum, meaning that the (real) velocity field is of less physical significance. Correspondingly, the fluidic approximation recovers less of the adiabatic density error, but nevertheless reduces it from $\approx 8\%$ to $\approx 4\%$, at $t=15$ a.u. Accordingly, the tunneling rate from the left-hand side to the right-hand side is initially improved, but this is not the case at later times [inset of Fig.~\ref{tunneling_1}].

In summary, we have calculated the \textit{exact} adiabatic and non-adiabatic parts of the KS potential, $v_{\mathrm{KS}}^{\mathrm{A}}$ and $\Delta v_{\mathrm{KS}}$, for a variety of model systems. $\Delta v_{\mathrm{KS}}$ is precisely defined by our procedure, and represents the part of the time-dependent KS potential that is \textit{intrinsically unobtainable from a ground-state functional}. Our key finding is that a simple geometrical approximation to this non-adiabatic KS potential -- making use of a Galilean transformation to the local instantaneous rest frame -- recovers most of the density error attributable to the exact adiabatic approach: typically $80-95 \%$ in the ballistic systems studied. Studies of additional systems should further illuminate this decomposition of the KS potential of TDDFT in highly non-adiabatic situations, with the fluidic approximation providing a solid foundation for a hierarchy of approximations to $\Delta v_{\mathrm{KS}}$.

We thank Jack Wetherell and Nick Woods for recent developments in the iDEA code, and Matt Hodgson and Carsten Ullrich for helpful comments. Data created during this research is available from the York Research Database \footnote{M. T. Entwistle and R. W. Godby, Data related to ``Exact non-adiabatic part of the Kohn-Sham potential and its fluidic approximation'', http://dx.doi.org/10.15124/8570b943-498b-4690-8044-b2208d318ef0 (2020).}.

\bibliographystyle{apsrev4-2}
\bibliography{Entwistle_2019}

\begin{thebibliography}{73}%
\makeatletter
\providecommand \@ifxundefined [1]{%
 \@ifx{#1\undefined}
}%
\providecommand \@ifnum [1]{%
 \ifnum #1\expandafter \@firstoftwo
 \else \expandafter \@secondoftwo
 \fi
}%
\providecommand \@ifx [1]{%
 \ifx #1\expandafter \@firstoftwo
 \else \expandafter \@secondoftwo
 \fi
}%
\providecommand \natexlab [1]{#1}%
\providecommand \enquote  [1]{``#1''}%
\providecommand \bibnamefont  [1]{#1}%
\providecommand \bibfnamefont [1]{#1}%
\providecommand \citenamefont [1]{#1}%
\providecommand \href@noop [0]{\@secondoftwo}%
\providecommand \href [0]{\begingroup \@sanitize@url \@href}%
\providecommand \@href[1]{\@@startlink{#1}\@@href}%
\providecommand \@@href[1]{\endgroup#1\@@endlink}%
\providecommand \@sanitize@url [0]{\catcode `\\12\catcode `\$12\catcode
  `\&12\catcode `\#12\catcode `\^12\catcode `\_12\catcode `\%12\relax}%
\providecommand \@@startlink[1]{}%
\providecommand \@@endlink[0]{}%
\providecommand \url  [0]{\begingroup\@sanitize@url \@url }%
\providecommand \@url [1]{\endgroup\@href {#1}{\urlprefix }}%
\providecommand \urlprefix  [0]{URL }%
\providecommand \Eprint [0]{\href }%
\providecommand \doibase [0]{https://doi.org/}%
\providecommand \selectlanguage [0]{\@gobble}%
\providecommand \bibinfo  [0]{\@secondoftwo}%
\providecommand \bibfield  [0]{\@secondoftwo}%
\providecommand \translation [1]{[#1]}%
\providecommand \BibitemOpen [0]{}%
\providecommand \bibitemStop [0]{}%
\providecommand \bibitemNoStop [0]{.\EOS\space}%
\providecommand \EOS [0]{\spacefactor3000\relax}%
\providecommand \BibitemShut  [1]{\csname bibitem#1\endcsname}%
\let\auto@bib@innerbib\@empty
\bibitem [{\citenamefont {Runge}\ and\ \citenamefont {Gross}(1984)}]{TDDFT}%
  \BibitemOpen
  \bibfield  {author} {\bibinfo {author} {\bibfnamefont {E.}~\bibnamefont
  {Runge}}\ and\ \bibinfo {author} {\bibfnamefont {E.~K.~U.}\ \bibnamefont
  {Gross}},\ }\href {https://doi.org/10.1103/PhysRevLett.52.997} {\bibfield
  {journal} {\bibinfo  {journal} {Phys. Rev. Lett.}\ }\textbf {\bibinfo
  {volume} {52}},\ \bibinfo {pages} {997} (\bibinfo {year} {1984})}\BibitemShut
  {NoStop}%
\bibitem [{\citenamefont {van Leeuwen}(1999)}]{TDDFT2}%
  \BibitemOpen
  \bibfield  {author} {\bibinfo {author} {\bibfnamefont {R.}~\bibnamefont {van
  Leeuwen}},\ }\href {https://doi.org/10.1103/PhysRevLett.82.3863} {\bibfield
  {journal} {\bibinfo  {journal} {Phys. Rev. Lett.}\ }\textbf {\bibinfo
  {volume} {82}},\ \bibinfo {pages} {3863} (\bibinfo {year}
  {1999})}\BibitemShut {NoStop}%
\bibitem [{\citenamefont {Kohn}\ and\ \citenamefont {Sham}(1965)}]{KS}%
  \BibitemOpen
  \bibfield  {author} {\bibinfo {author} {\bibfnamefont {W.}~\bibnamefont
  {Kohn}}\ and\ \bibinfo {author} {\bibfnamefont {L.~J.}\ \bibnamefont
  {Sham}},\ }\href {https://doi.org/10.1103/PhysRev.140.A1133} {\bibfield
  {journal} {\bibinfo  {journal} {Phys. Rev.}\ }\textbf {\bibinfo {volume}
  {140}},\ \bibinfo {pages} {A1133} (\bibinfo {year} {1965})}\BibitemShut
  {NoStop}%
\bibitem [{\citenamefont {Perdew}\ and\ \citenamefont {Zunger}(1981)}]{PZ_LDA}%
  \BibitemOpen
  \bibfield  {author} {\bibinfo {author} {\bibfnamefont {J.~P.}\ \bibnamefont
  {Perdew}}\ and\ \bibinfo {author} {\bibfnamefont {A.}~\bibnamefont
  {Zunger}},\ }\href {https://doi.org/10.1103/PhysRevB.23.5048} {\bibfield
  {journal} {\bibinfo  {journal} {Phys. Rev. B}\ }\textbf {\bibinfo {volume}
  {23}},\ \bibinfo {pages} {5048} (\bibinfo {year} {1981})}\BibitemShut
  {NoStop}%
\bibitem [{\citenamefont {Perdew}\ and\ \citenamefont {Wang}(1992)}]{PW_LDA}%
  \BibitemOpen
  \bibfield  {author} {\bibinfo {author} {\bibfnamefont {J.~P.}\ \bibnamefont
  {Perdew}}\ and\ \bibinfo {author} {\bibfnamefont {Y.}~\bibnamefont {Wang}},\
  }\href {https://doi.org/10.1103/PhysRevB.45.13244} {\bibfield  {journal}
  {\bibinfo  {journal} {Phys. Rev. B}\ }\textbf {\bibinfo {volume} {45}},\
  \bibinfo {pages} {13244} (\bibinfo {year} {1992})}\BibitemShut {NoStop}%
\bibitem [{\citenamefont {Perdew}(1985)}]{GGA}%
  \BibitemOpen
  \bibfield  {author} {\bibinfo {author} {\bibfnamefont {J.~P.}\ \bibnamefont
  {Perdew}},\ }\href {https://doi.org/10.1103/PhysRevLett.55.1665} {\bibfield
  {journal} {\bibinfo  {journal} {Phys. Rev. Lett.}\ }\textbf {\bibinfo
  {volume} {55}},\ \bibinfo {pages} {1665} (\bibinfo {year}
  {1985})}\BibitemShut {NoStop}%
\bibitem [{\citenamefont {Burke}\ \emph {et~al.}(1997)\citenamefont {Burke},
  \citenamefont {Perdew},\ and\ \citenamefont {Wang}}]{PW91}%
  \BibitemOpen
  \bibfield  {author} {\bibinfo {author} {\bibfnamefont {K.}~\bibnamefont
  {Burke}}, \bibinfo {author} {\bibfnamefont {J.~P.}\ \bibnamefont {Perdew}},\
  and\ \bibinfo {author} {\bibfnamefont {Y.}~\bibnamefont {Wang}},\ }\bibinfo
  {title} {Derivation of a generalized gradient approximation: The pw91 density
  functional},\ in\ \href
  {http://link.springer.com/chapter/10.1007%2F978-1-4899-0316-7_7} {\emph
  {\bibinfo {booktitle} {Electronic Density Functional Theory: Recent Progress
  and New Directions}}},\ \bibinfo {editor} {edited by\ \bibinfo {editor}
  {\bibfnamefont {J.~F.}\ \bibnamefont {Dobson}}, \bibinfo {editor}
  {\bibfnamefont {G.}~\bibnamefont {Vignale}},\ and\ \bibinfo {editor}
  {\bibfnamefont {M.~P.}\ \bibnamefont {Das}}}\ (\bibinfo  {publisher}
  {Plenum},\ \bibinfo {address} {NY},\ \bibinfo {year} {1997})\ p.~\bibinfo
  {pages} {81}\BibitemShut {NoStop}%
\bibitem [{\citenamefont {Perdew}\ \emph {et~al.}(1992)\citenamefont {Perdew},
  \citenamefont {Chevary}, \citenamefont {Vosko}, \citenamefont {Jackson},
  \citenamefont {Pederson}, \citenamefont {Singh},\ and\ \citenamefont
  {Fiolhais}}]{PW_GGA}%
  \BibitemOpen
  \bibfield  {author} {\bibinfo {author} {\bibfnamefont {J.~P.}\ \bibnamefont
  {Perdew}}, \bibinfo {author} {\bibfnamefont {J.~A.}\ \bibnamefont {Chevary}},
  \bibinfo {author} {\bibfnamefont {S.~H.}\ \bibnamefont {Vosko}}, \bibinfo
  {author} {\bibfnamefont {K.~A.}\ \bibnamefont {Jackson}}, \bibinfo {author}
  {\bibfnamefont {M.~R.}\ \bibnamefont {Pederson}}, \bibinfo {author}
  {\bibfnamefont {D.~J.}\ \bibnamefont {Singh}},\ and\ \bibinfo {author}
  {\bibfnamefont {C.}~\bibnamefont {Fiolhais}},\ }\href
  {https://doi.org/10.1103/PhysRevB.46.6671} {\bibfield  {journal} {\bibinfo
  {journal} {Phys. Rev. B}\ }\textbf {\bibinfo {volume} {46}},\ \bibinfo
  {pages} {6671} (\bibinfo {year} {1992})}\BibitemShut {NoStop}%
\bibitem [{\citenamefont {Perdew}\ \emph
  {et~al.}(1996{\natexlab{a}})\citenamefont {Perdew}, \citenamefont {Burke},\
  and\ \citenamefont {Ernzerhof}}]{GGA2}%
  \BibitemOpen
  \bibfield  {author} {\bibinfo {author} {\bibfnamefont {J.~P.}\ \bibnamefont
  {Perdew}}, \bibinfo {author} {\bibfnamefont {K.}~\bibnamefont {Burke}},\ and\
  \bibinfo {author} {\bibfnamefont {M.}~\bibnamefont {Ernzerhof}},\ }\href
  {https://doi.org/10.1103/PhysRevLett.77.3865} {\bibfield  {journal} {\bibinfo
   {journal} {Phys. Rev. Lett.}\ }\textbf {\bibinfo {volume} {77}},\ \bibinfo
  {pages} {3865} (\bibinfo {year} {1996}{\natexlab{a}})}\BibitemShut {NoStop}%
\bibitem [{\citenamefont {Perdew}\ \emph
  {et~al.}(1996{\natexlab{b}})\citenamefont {Perdew}, \citenamefont
  {Ernzerhof},\ and\ \citenamefont {Burke}}]{Hybrid_GGA}%
  \BibitemOpen
  \bibfield  {author} {\bibinfo {author} {\bibfnamefont {J.~P.}\ \bibnamefont
  {Perdew}}, \bibinfo {author} {\bibfnamefont {M.}~\bibnamefont {Ernzerhof}},\
  and\ \bibinfo {author} {\bibfnamefont {K.}~\bibnamefont {Burke}},\ }\href
  {https://doi.org/10.1063/1.472933} {\bibfield  {journal} {\bibinfo  {journal}
  {The Journal of Chemical Physics}\ }\textbf {\bibinfo {volume} {105}},\
  \bibinfo {pages} {9982} (\bibinfo {year} {1996}{\natexlab{b}})},\ \Eprint
  {https://arxiv.org/abs/https://doi.org/10.1063/1.472933}
  {https://doi.org/10.1063/1.472933} \BibitemShut {NoStop}%
\bibitem [{\citenamefont {Proynov}\ \emph {et~al.}(1997)\citenamefont
  {Proynov}, \citenamefont {Sirois},\ and\ \citenamefont {Salahub}}]{Meta_GGA}%
  \BibitemOpen
  \bibfield  {author} {\bibinfo {author} {\bibfnamefont {E.}~\bibnamefont
  {Proynov}}, \bibinfo {author} {\bibfnamefont {S.}~\bibnamefont {Sirois}},\
  and\ \bibinfo {author} {\bibfnamefont {D.}~\bibnamefont {Salahub}},\ }\href
  {https://doi.org/10.1002/(SICI)1097-461X(1997)64:4<427::AID-QUA5>3.0.CO;2-Y}
  {\bibfield  {journal} {\bibinfo  {journal} {International Journal of Quantum
  Chemistry}\ }\textbf {\bibinfo {volume} {64}},\ \bibinfo {pages} {427 }
  (\bibinfo {year} {1997})}\BibitemShut {NoStop}%
\bibitem [{\citenamefont {Van~Voorhis}\ and\ \citenamefont
  {Scuseria}(1998)}]{Meta_GGA2}%
  \BibitemOpen
  \bibfield  {author} {\bibinfo {author} {\bibfnamefont {T.}~\bibnamefont
  {Van~Voorhis}}\ and\ \bibinfo {author} {\bibfnamefont {G.~E.}\ \bibnamefont
  {Scuseria}},\ }\href {https://doi.org/10.1063/1.476577} {\bibfield  {journal}
  {\bibinfo  {journal} {The Journal of Chemical Physics}\ }\textbf {\bibinfo
  {volume} {109}},\ \bibinfo {pages} {400} (\bibinfo {year} {1998})},\ \Eprint
  {https://arxiv.org/abs/https://doi.org/10.1063/1.476577}
  {https://doi.org/10.1063/1.476577} \BibitemShut {NoStop}%
\bibitem [{\citenamefont {Perdew}\ \emph {et~al.}(1999)\citenamefont {Perdew},
  \citenamefont {Kurth}, \citenamefont {Zupan},\ and\ \citenamefont
  {Blaha}}]{Meta_GGA3}%
  \BibitemOpen
  \bibfield  {author} {\bibinfo {author} {\bibfnamefont {J.~P.}\ \bibnamefont
  {Perdew}}, \bibinfo {author} {\bibfnamefont {S.}~\bibnamefont {Kurth}},
  \bibinfo {author} {\bibfnamefont {A.~c.~v.}\ \bibnamefont {Zupan}},\ and\
  \bibinfo {author} {\bibfnamefont {P.}~\bibnamefont {Blaha}},\ }\href
  {https://doi.org/10.1103/PhysRevLett.82.2544} {\bibfield  {journal} {\bibinfo
   {journal} {Phys. Rev. Lett.}\ }\textbf {\bibinfo {volume} {82}},\ \bibinfo
  {pages} {2544} (\bibinfo {year} {1999})}\BibitemShut {NoStop}%
\bibitem [{\citenamefont {Sun}\ \emph {et~al.}(2015)\citenamefont {Sun},
  \citenamefont {Ruzsinszky},\ and\ \citenamefont {Perdew}}]{SCAN}%
  \BibitemOpen
  \bibfield  {author} {\bibinfo {author} {\bibfnamefont {J.}~\bibnamefont
  {Sun}}, \bibinfo {author} {\bibfnamefont {A.}~\bibnamefont {Ruzsinszky}},\
  and\ \bibinfo {author} {\bibfnamefont {J.~P.}\ \bibnamefont {Perdew}},\
  }\href {https://doi.org/10.1103/PhysRevLett.115.036402} {\bibfield  {journal}
  {\bibinfo  {journal} {Phys. Rev. Lett.}\ }\textbf {\bibinfo {volume} {115}},\
  \bibinfo {pages} {036402} (\bibinfo {year} {2015})}\BibitemShut {NoStop}%
\bibitem [{\citenamefont {Maitra}(2017)}]{Adiabatic_fail10}%
  \BibitemOpen
  \bibfield  {author} {\bibinfo {author} {\bibfnamefont {N.~T.}\ \bibnamefont
  {Maitra}},\ }\href {https://doi.org/10.1088/1361-648x/aa836e} {\bibfield
  {journal} {\bibinfo  {journal} {Journal of Physics: Condensed Matter}\
  }\textbf {\bibinfo {volume} {29}},\ \bibinfo {pages} {423001} (\bibinfo
  {year} {2017})}\BibitemShut {NoStop}%
\bibitem [{\citenamefont {Elliott}\ \emph {et~al.}(2012)\citenamefont
  {Elliott}, \citenamefont {Fuks}, \citenamefont {Rubio},\ and\ \citenamefont
  {Maitra}}]{Elliot_2012}%
  \BibitemOpen
  \bibfield  {author} {\bibinfo {author} {\bibfnamefont {P.}~\bibnamefont
  {Elliott}}, \bibinfo {author} {\bibfnamefont {J.~I.}\ \bibnamefont {Fuks}},
  \bibinfo {author} {\bibfnamefont {A.}~\bibnamefont {Rubio}},\ and\ \bibinfo
  {author} {\bibfnamefont {N.~T.}\ \bibnamefont {Maitra}},\ }\href
  {https://doi.org/10.1103/PhysRevLett.109.266404} {\bibfield  {journal}
  {\bibinfo  {journal} {Phys. Rev. Lett.}\ }\textbf {\bibinfo {volume} {109}},\
  \bibinfo {pages} {266404} (\bibinfo {year} {2012})}\BibitemShut {NoStop}%
\bibitem [{\citenamefont {Fuks}(2016)}]{Adiabatic_fail9}%
  \BibitemOpen
  \bibfield  {author} {\bibinfo {author} {\bibfnamefont {J.~I.}\ \bibnamefont
  {Fuks}},\ }\href@noop {} {\bibfield  {journal} {\bibinfo  {journal} {The
  European Physical Journal B}\ }\textbf {\bibinfo {volume} {89}},\ \bibinfo
  {pages} {236} (\bibinfo {year} {2016})}\BibitemShut {NoStop}%
\bibitem [{\citenamefont {Fuks}\ \emph {et~al.}(2015)\citenamefont {Fuks},
  \citenamefont {Luo}, \citenamefont {Sandoval},\ and\ \citenamefont
  {Maitra}}]{TR_spectroscopy}%
  \BibitemOpen
  \bibfield  {author} {\bibinfo {author} {\bibfnamefont {J.~I.}\ \bibnamefont
  {Fuks}}, \bibinfo {author} {\bibfnamefont {K.}~\bibnamefont {Luo}}, \bibinfo
  {author} {\bibfnamefont {E.~D.}\ \bibnamefont {Sandoval}},\ and\ \bibinfo
  {author} {\bibfnamefont {N.~T.}\ \bibnamefont {Maitra}},\ }\href
  {https://doi.org/10.1103/PhysRevLett.114.183002} {\bibfield  {journal}
  {\bibinfo  {journal} {Phys. Rev. Lett.}\ }\textbf {\bibinfo {volume} {114}},\
  \bibinfo {pages} {183002} (\bibinfo {year} {2015})}\BibitemShut {NoStop}%
\bibitem [{\citenamefont {Suzuki}\ \emph {et~al.}(2017)\citenamefont {Suzuki},
  \citenamefont {Lacombe}, \citenamefont {Watanabe},\ and\ \citenamefont
  {Maitra}}]{Electron_scattering}%
  \BibitemOpen
  \bibfield  {author} {\bibinfo {author} {\bibfnamefont {Y.}~\bibnamefont
  {Suzuki}}, \bibinfo {author} {\bibfnamefont {L.}~\bibnamefont {Lacombe}},
  \bibinfo {author} {\bibfnamefont {K.}~\bibnamefont {Watanabe}},\ and\
  \bibinfo {author} {\bibfnamefont {N.~T.}\ \bibnamefont {Maitra}},\ }\href
  {https://doi.org/10.1103/PhysRevLett.119.263401} {\bibfield  {journal}
  {\bibinfo  {journal} {Phys. Rev. Lett.}\ }\textbf {\bibinfo {volume} {119}},\
  \bibinfo {pages} {263401} (\bibinfo {year} {2017})}\BibitemShut {NoStop}%
\bibitem [{Note1()}]{Note1}%
  \BibitemOpen
  \bibinfo {note} {The real system of interacting electrons is mapped onto an
  auxiliary system of noninteracting electrons moving in the effective
  potential $v_{\protect \mathrm {KS}}$.}\BibitemShut {Stop}%
\bibitem [{\citenamefont {Dobson}(1994)}]{HPT}%
  \BibitemOpen
  \bibfield  {author} {\bibinfo {author} {\bibfnamefont {J.~F.}\ \bibnamefont
  {Dobson}},\ }\href {https://doi.org/10.1103/PhysRevLett.73.2244} {\bibfield
  {journal} {\bibinfo  {journal} {Phys. Rev. Lett.}\ }\textbf {\bibinfo
  {volume} {73}},\ \bibinfo {pages} {2244} (\bibinfo {year}
  {1994})}\BibitemShut {NoStop}%
\bibitem [{\citenamefont {Vignale}(1995{\natexlab{a}})}]{zeroforce}%
  \BibitemOpen
  \bibfield  {author} {\bibinfo {author} {\bibfnamefont {G.}~\bibnamefont
  {Vignale}},\ }\href {https://doi.org/10.1103/PhysRevLett.74.3233} {\bibfield
  {journal} {\bibinfo  {journal} {Phys. Rev. Lett.}\ }\textbf {\bibinfo
  {volume} {74}},\ \bibinfo {pages} {3233} (\bibinfo {year}
  {1995}{\natexlab{a}})}\BibitemShut {NoStop}%
\bibitem [{\citenamefont {Vignale}(1995{\natexlab{b}})}]{fxc_sumrule}%
  \BibitemOpen
  \bibfield  {author} {\bibinfo {author} {\bibfnamefont {G.}~\bibnamefont
  {Vignale}},\ }\href
  {https://doi.org/https://doi.org/10.1016/0375-9601(95)00855-3} {\bibfield
  {journal} {\bibinfo  {journal} {Physics Letters A}\ }\textbf {\bibinfo
  {volume} {209}},\ \bibinfo {pages} {206 } (\bibinfo {year}
  {1995}{\natexlab{b}})}\BibitemShut {NoStop}%
\bibitem [{\citenamefont {Vignale}\ and\ \citenamefont {Kohn}(1996)}]{VK}%
  \BibitemOpen
  \bibfield  {author} {\bibinfo {author} {\bibfnamefont {G.}~\bibnamefont
  {Vignale}}\ and\ \bibinfo {author} {\bibfnamefont {W.}~\bibnamefont {Kohn}},\
  }\href {https://doi.org/10.1103/PhysRevLett.77.2037} {\bibfield  {journal}
  {\bibinfo  {journal} {Phys. Rev. Lett.}\ }\textbf {\bibinfo {volume} {77}},\
  \bibinfo {pages} {2037} (\bibinfo {year} {1996})}\BibitemShut {NoStop}%
\bibitem [{\citenamefont {Jamorski}\ \emph {et~al.}(1996)\citenamefont
  {Jamorski}, \citenamefont {Casida},\ and\ \citenamefont
  {Salahub}}]{Adiabatic_fail}%
  \BibitemOpen
  \bibfield  {author} {\bibinfo {author} {\bibfnamefont {C.}~\bibnamefont
  {Jamorski}}, \bibinfo {author} {\bibfnamefont {M.~E.}\ \bibnamefont
  {Casida}},\ and\ \bibinfo {author} {\bibfnamefont {D.~R.}\ \bibnamefont
  {Salahub}},\ }\href {https://doi.org/10.1063/1.471140} {\bibfield  {journal}
  {\bibinfo  {journal} {The Journal of Chemical Physics}\ }\textbf {\bibinfo
  {volume} {104}},\ \bibinfo {pages} {5134} (\bibinfo {year} {1996})},\ \Eprint
  {https://arxiv.org/abs/https://doi.org/10.1063/1.471140}
  {https://doi.org/10.1063/1.471140} \BibitemShut {NoStop}%
\bibitem [{\citenamefont {Hessler}\ \emph {et~al.}(2002)\citenamefont
  {Hessler}, \citenamefont {Maitra},\ and\ \citenamefont
  {Burke}}]{Hessler_2002}%
  \BibitemOpen
  \bibfield  {author} {\bibinfo {author} {\bibfnamefont {P.}~\bibnamefont
  {Hessler}}, \bibinfo {author} {\bibfnamefont {N.~T.}\ \bibnamefont
  {Maitra}},\ and\ \bibinfo {author} {\bibfnamefont {K.}~\bibnamefont
  {Burke}},\ }\href {https://doi.org/10.1063/1.1479349} {\bibfield  {journal}
  {\bibinfo  {journal} {The Journal of Chemical Physics}\ }\textbf {\bibinfo
  {volume} {117}},\ \bibinfo {pages} {72} (\bibinfo {year} {2002})},\ \Eprint
  {https://arxiv.org/abs/https://doi.org/10.1063/1.1479349}
  {https://doi.org/10.1063/1.1479349} \BibitemShut {NoStop}%
\bibitem [{\citenamefont {Dreuw}\ \emph {et~al.}(2003)\citenamefont {Dreuw},
  \citenamefont {Weisman},\ and\ \citenamefont
  {Head-Gordon}}]{Adiabatic_fail2}%
  \BibitemOpen
  \bibfield  {author} {\bibinfo {author} {\bibfnamefont {A.}~\bibnamefont
  {Dreuw}}, \bibinfo {author} {\bibfnamefont {J.~L.}\ \bibnamefont {Weisman}},\
  and\ \bibinfo {author} {\bibfnamefont {M.}~\bibnamefont {Head-Gordon}},\
  }\href {https://doi.org/10.1063/1.1590951} {\bibfield  {journal} {\bibinfo
  {journal} {The Journal of Chemical Physics}\ }\textbf {\bibinfo {volume}
  {119}},\ \bibinfo {pages} {2943} (\bibinfo {year} {2003})},\ \Eprint
  {https://arxiv.org/abs/https://doi.org/10.1063/1.1590951}
  {https://doi.org/10.1063/1.1590951} \BibitemShut {NoStop}%
\bibitem [{\citenamefont {Maitra}\ \emph {et~al.}(2004)\citenamefont {Maitra},
  \citenamefont {Zhang}, \citenamefont {Cave},\ and\ \citenamefont
  {Burke}}]{Adiabatic_fail3}%
  \BibitemOpen
  \bibfield  {author} {\bibinfo {author} {\bibfnamefont {N.~T.}\ \bibnamefont
  {Maitra}}, \bibinfo {author} {\bibfnamefont {F.}~\bibnamefont {Zhang}},
  \bibinfo {author} {\bibfnamefont {R.~J.}\ \bibnamefont {Cave}},\ and\
  \bibinfo {author} {\bibfnamefont {K.}~\bibnamefont {Burke}},\ }\href
  {https://doi.org/10.1063/1.1651060} {\bibfield  {journal} {\bibinfo
  {journal} {The Journal of Chemical Physics}\ }\textbf {\bibinfo {volume}
  {120}},\ \bibinfo {pages} {5932} (\bibinfo {year} {2004})},\ \Eprint
  {https://arxiv.org/abs/https://doi.org/10.1063/1.1651060}
  {https://doi.org/10.1063/1.1651060} \BibitemShut {NoStop}%
\bibitem [{\citenamefont {Neugebauer}\ \emph {et~al.}(2004)\citenamefont
  {Neugebauer}, \citenamefont {Baerends},\ and\ \citenamefont
  {Nooijen}}]{Adiabatic_fail4}%
  \BibitemOpen
  \bibfield  {author} {\bibinfo {author} {\bibfnamefont {J.}~\bibnamefont
  {Neugebauer}}, \bibinfo {author} {\bibfnamefont {E.~J.}\ \bibnamefont
  {Baerends}},\ and\ \bibinfo {author} {\bibfnamefont {M.}~\bibnamefont
  {Nooijen}},\ }\href {https://doi.org/10.1063/1.1785775} {\bibfield  {journal}
  {\bibinfo  {journal} {The Journal of Chemical Physics}\ }\textbf {\bibinfo
  {volume} {121}},\ \bibinfo {pages} {6155} (\bibinfo {year} {2004})},\ \Eprint
  {https://arxiv.org/abs/https://doi.org/10.1063/1.1785775}
  {https://doi.org/10.1063/1.1785775} \BibitemShut {NoStop}%
\bibitem [{\citenamefont {Cave}\ \emph {et~al.}(2004)\citenamefont {Cave},
  \citenamefont {Zhang}, \citenamefont {Maitra},\ and\ \citenamefont
  {Burke}}]{Adiabatic_fail5}%
  \BibitemOpen
  \bibfield  {author} {\bibinfo {author} {\bibfnamefont {R.~J.}\ \bibnamefont
  {Cave}}, \bibinfo {author} {\bibfnamefont {F.}~\bibnamefont {Zhang}},
  \bibinfo {author} {\bibfnamefont {N.~T.}\ \bibnamefont {Maitra}},\ and\
  \bibinfo {author} {\bibfnamefont {K.}~\bibnamefont {Burke}},\ }\href
  {https://doi.org/https://doi.org/10.1016/j.cplett.2004.03.051} {\bibfield
  {journal} {\bibinfo  {journal} {Chemical Physics Letters}\ }\textbf {\bibinfo
  {volume} {389}},\ \bibinfo {pages} {39 } (\bibinfo {year}
  {2004})}\BibitemShut {NoStop}%
\bibitem [{\citenamefont {Ullrich}\ and\ \citenamefont
  {Tokatly}(2006)}]{Adiabatic_fail6}%
  \BibitemOpen
  \bibfield  {author} {\bibinfo {author} {\bibfnamefont {C.~A.}\ \bibnamefont
  {Ullrich}}\ and\ \bibinfo {author} {\bibfnamefont {I.~V.}\ \bibnamefont
  {Tokatly}},\ }\href {https://doi.org/10.1103/PhysRevB.73.235102} {\bibfield
  {journal} {\bibinfo  {journal} {Phys. Rev. B}\ }\textbf {\bibinfo {volume}
  {73}},\ \bibinfo {pages} {235102} (\bibinfo {year} {2006})}\BibitemShut
  {NoStop}%
\bibitem [{\citenamefont {Giesbertz}\ \emph {et~al.}(2008)\citenamefont
  {Giesbertz}, \citenamefont {Baerends},\ and\ \citenamefont
  {Gritsenko}}]{Adiabatic_fail7}%
  \BibitemOpen
  \bibfield  {author} {\bibinfo {author} {\bibfnamefont {K.~J.~H.}\
  \bibnamefont {Giesbertz}}, \bibinfo {author} {\bibfnamefont {E.~J.}\
  \bibnamefont {Baerends}},\ and\ \bibinfo {author} {\bibfnamefont {O.~V.}\
  \bibnamefont {Gritsenko}},\ }\href
  {https://doi.org/10.1103/PhysRevLett.101.033004} {\bibfield  {journal}
  {\bibinfo  {journal} {Phys. Rev. Lett.}\ }\textbf {\bibinfo {volume} {101}},\
  \bibinfo {pages} {033004} (\bibinfo {year} {2008})}\BibitemShut {NoStop}%
\bibitem [{\citenamefont {Fuks}\ \emph {et~al.}(2011)\citenamefont {Fuks},
  \citenamefont {Rubio},\ and\ \citenamefont {Maitra}}]{Adiabatic_fail8}%
  \BibitemOpen
  \bibfield  {author} {\bibinfo {author} {\bibfnamefont {J.~I.}\ \bibnamefont
  {Fuks}}, \bibinfo {author} {\bibfnamefont {A.}~\bibnamefont {Rubio}},\ and\
  \bibinfo {author} {\bibfnamefont {N.~T.}\ \bibnamefont {Maitra}},\ }\href
  {https://doi.org/10.1103/PhysRevA.83.042501} {\bibfield  {journal} {\bibinfo
  {journal} {Phys. Rev. A}\ }\textbf {\bibinfo {volume} {83}},\ \bibinfo
  {pages} {042501} (\bibinfo {year} {2011})}\BibitemShut {NoStop}%
\bibitem [{\citenamefont {Maitra}(2016)}]{Maitra_2016}%
  \BibitemOpen
  \bibfield  {author} {\bibinfo {author} {\bibfnamefont {N.~T.}\ \bibnamefont
  {Maitra}},\ }\href {https://doi.org/10.1063/1.4953039} {\bibfield  {journal}
  {\bibinfo  {journal} {The Journal of Chemical Physics}\ }\textbf {\bibinfo
  {volume} {144}},\ \bibinfo {pages} {220901} (\bibinfo {year} {2016})},\
  \Eprint {https://arxiv.org/abs/https://doi.org/10.1063/1.4953039}
  {https://doi.org/10.1063/1.4953039} \BibitemShut {NoStop}%
\bibitem [{\citenamefont {Singh}\ \emph {et~al.}(2019)\citenamefont {Singh},
  \citenamefont {Elliott}, \citenamefont {Nautiyal}, \citenamefont {Dewhurst},\
  and\ \citenamefont {Sharma}}]{Adiabatic_fail11}%
  \BibitemOpen
  \bibfield  {author} {\bibinfo {author} {\bibfnamefont {N.}~\bibnamefont
  {Singh}}, \bibinfo {author} {\bibfnamefont {P.}~\bibnamefont {Elliott}},
  \bibinfo {author} {\bibfnamefont {T.}~\bibnamefont {Nautiyal}}, \bibinfo
  {author} {\bibfnamefont {J.~K.}\ \bibnamefont {Dewhurst}},\ and\ \bibinfo
  {author} {\bibfnamefont {S.}~\bibnamefont {Sharma}},\ }\href
  {https://doi.org/10.1103/PhysRevB.99.035151} {\bibfield  {journal} {\bibinfo
  {journal} {Phys. Rev. B}\ }\textbf {\bibinfo {volume} {99}},\ \bibinfo
  {pages} {035151} (\bibinfo {year} {2019})}\BibitemShut {NoStop}%
\bibitem [{\citenamefont {Gross}\ and\ \citenamefont {Kohn}(1985)}]{GK}%
  \BibitemOpen
  \bibfield  {author} {\bibinfo {author} {\bibfnamefont {E.~K.~U.}\
  \bibnamefont {Gross}}\ and\ \bibinfo {author} {\bibfnamefont
  {W.}~\bibnamefont {Kohn}},\ }\href
  {https://doi.org/10.1103/PhysRevLett.55.2850} {\bibfield  {journal} {\bibinfo
   {journal} {Phys. Rev. Lett.}\ }\textbf {\bibinfo {volume} {55}},\ \bibinfo
  {pages} {2850} (\bibinfo {year} {1985})}\BibitemShut {NoStop}%
\bibitem [{\citenamefont {Dobson}\ \emph {et~al.}(2013)\citenamefont {Dobson},
  \citenamefont {Vignale},\ and\ \citenamefont {Das}}]{VK2}%
  \BibitemOpen
  \bibfield  {author} {\bibinfo {author} {\bibfnamefont {J.}~\bibnamefont
  {Dobson}}, \bibinfo {author} {\bibfnamefont {G.}~\bibnamefont {Vignale}},\
  and\ \bibinfo {author} {\bibfnamefont {M.}~\bibnamefont {Das}},\ }\href
  {https://books.google.co.uk/books?id=4OvvBwAAQBAJ} {\emph {\bibinfo {title}
  {Electronic Density Functional Theory: Recent Progress and New Directions}}}\
  (\bibinfo  {publisher} {Springer US},\ \bibinfo {year} {2013})\BibitemShut
  {NoStop}%
\bibitem [{\citenamefont {Vignale}\ \emph {et~al.}(1997)\citenamefont
  {Vignale}, \citenamefont {Ullrich},\ and\ \citenamefont {Conti}}]{VK_fluid}%
  \BibitemOpen
  \bibfield  {author} {\bibinfo {author} {\bibfnamefont {G.}~\bibnamefont
  {Vignale}}, \bibinfo {author} {\bibfnamefont {C.~A.}\ \bibnamefont
  {Ullrich}},\ and\ \bibinfo {author} {\bibfnamefont {S.}~\bibnamefont
  {Conti}},\ }\href {https://doi.org/10.1103/PhysRevLett.79.4878} {\bibfield
  {journal} {\bibinfo  {journal} {Phys. Rev. Lett.}\ }\textbf {\bibinfo
  {volume} {79}},\ \bibinfo {pages} {4878} (\bibinfo {year}
  {1997})}\BibitemShut {NoStop}%
\bibitem [{\citenamefont {van Faassen}\ \emph {et~al.}(2002)\citenamefont {van
  Faassen}, \citenamefont {de~Boeij}, \citenamefont {van Leeuwen},
  \citenamefont {Berger},\ and\ \citenamefont {Snijders}}]{VK_good}%
  \BibitemOpen
  \bibfield  {author} {\bibinfo {author} {\bibfnamefont {M.}~\bibnamefont {van
  Faassen}}, \bibinfo {author} {\bibfnamefont {P.~L.}\ \bibnamefont
  {de~Boeij}}, \bibinfo {author} {\bibfnamefont {R.}~\bibnamefont {van
  Leeuwen}}, \bibinfo {author} {\bibfnamefont {J.~A.}\ \bibnamefont {Berger}},\
  and\ \bibinfo {author} {\bibfnamefont {J.~G.}\ \bibnamefont {Snijders}},\
  }\href {https://doi.org/10.1103/PhysRevLett.88.186401} {\bibfield  {journal}
  {\bibinfo  {journal} {Phys. Rev. Lett.}\ }\textbf {\bibinfo {volume} {88}},\
  \bibinfo {pages} {186401} (\bibinfo {year} {2002})}\BibitemShut {NoStop}%
\bibitem [{\citenamefont {van Faassen}\ \emph {et~al.}(2003)\citenamefont {van
  Faassen}, \citenamefont {de~Boeij}, \citenamefont {van Leeuwen},
  \citenamefont {Berger},\ and\ \citenamefont {Snijders}}]{VK_good2}%
  \BibitemOpen
  \bibfield  {author} {\bibinfo {author} {\bibfnamefont {M.}~\bibnamefont {van
  Faassen}}, \bibinfo {author} {\bibfnamefont {P.~L.}\ \bibnamefont
  {de~Boeij}}, \bibinfo {author} {\bibfnamefont {R.}~\bibnamefont {van
  Leeuwen}}, \bibinfo {author} {\bibfnamefont {J.~A.}\ \bibnamefont {Berger}},\
  and\ \bibinfo {author} {\bibfnamefont {J.~G.}\ \bibnamefont {Snijders}},\
  }\href {https://doi.org/10.1063/1.1529679} {\bibfield  {journal} {\bibinfo
  {journal} {The Journal of Chemical Physics}\ }\textbf {\bibinfo {volume}
  {118}},\ \bibinfo {pages} {1044} (\bibinfo {year} {2003})},\ \Eprint
  {https://arxiv.org/abs/https://doi.org/10.1063/1.1529679}
  {https://doi.org/10.1063/1.1529679} \BibitemShut {NoStop}%
\bibitem [{\citenamefont {de~Boeij}\ \emph {et~al.}(2001)\citenamefont
  {de~Boeij}, \citenamefont {Kootstra}, \citenamefont {Berger}, \citenamefont
  {van Leeuwen},\ and\ \citenamefont {Snijders}}]{VK_good3}%
  \BibitemOpen
  \bibfield  {author} {\bibinfo {author} {\bibfnamefont {P.~L.}\ \bibnamefont
  {de~Boeij}}, \bibinfo {author} {\bibfnamefont {F.}~\bibnamefont {Kootstra}},
  \bibinfo {author} {\bibfnamefont {J.~A.}\ \bibnamefont {Berger}}, \bibinfo
  {author} {\bibfnamefont {R.}~\bibnamefont {van Leeuwen}},\ and\ \bibinfo
  {author} {\bibfnamefont {J.~G.}\ \bibnamefont {Snijders}},\ }\href
  {https://doi.org/10.1063/1.1385370} {\bibfield  {journal} {\bibinfo
  {journal} {The Journal of Chemical Physics}\ }\textbf {\bibinfo {volume}
  {115}},\ \bibinfo {pages} {1995} (\bibinfo {year} {2001})},\ \Eprint
  {https://arxiv.org/abs/https://doi.org/10.1063/1.1385370}
  {https://doi.org/10.1063/1.1385370} \BibitemShut {NoStop}%
\bibitem [{\citenamefont {Ullrich}\ and\ \citenamefont
  {Vignale}(1998)}]{VK_good4}%
  \BibitemOpen
  \bibfield  {author} {\bibinfo {author} {\bibfnamefont {C.~A.}\ \bibnamefont
  {Ullrich}}\ and\ \bibinfo {author} {\bibfnamefont {G.}~\bibnamefont
  {Vignale}},\ }\href {https://doi.org/10.1103/PhysRevB.58.7141} {\bibfield
  {journal} {\bibinfo  {journal} {Phys. Rev. B}\ }\textbf {\bibinfo {volume}
  {58}},\ \bibinfo {pages} {7141} (\bibinfo {year} {1998})}\BibitemShut
  {NoStop}%
\bibitem [{\citenamefont {Ullrich}\ and\ \citenamefont
  {Vignale}(2001)}]{VK_good5}%
  \BibitemOpen
  \bibfield  {author} {\bibinfo {author} {\bibfnamefont {C.~A.}\ \bibnamefont
  {Ullrich}}\ and\ \bibinfo {author} {\bibfnamefont {G.}~\bibnamefont
  {Vignale}},\ }\href {https://doi.org/10.1103/PhysRevLett.87.037402}
  {\bibfield  {journal} {\bibinfo  {journal} {Phys. Rev. Lett.}\ }\textbf
  {\bibinfo {volume} {87}},\ \bibinfo {pages} {037402} (\bibinfo {year}
  {2001})}\BibitemShut {NoStop}%
\bibitem [{\citenamefont {Ullrich}\ and\ \citenamefont
  {Vignale}(2002)}]{VK_good6}%
  \BibitemOpen
  \bibfield  {author} {\bibinfo {author} {\bibfnamefont {C.~A.}\ \bibnamefont
  {Ullrich}}\ and\ \bibinfo {author} {\bibfnamefont {G.}~\bibnamefont
  {Vignale}},\ }\href {https://doi.org/10.1103/PhysRevB.65.245102} {\bibfield
  {journal} {\bibinfo  {journal} {Phys. Rev. B}\ }\textbf {\bibinfo {volume}
  {65}},\ \bibinfo {pages} {245102} (\bibinfo {year} {2002})}\BibitemShut
  {NoStop}%
\bibitem [{\citenamefont {Berger}\ \emph {et~al.}(2006)\citenamefont {Berger},
  \citenamefont {Romaniello}, \citenamefont {van Leeuwen},\ and\ \citenamefont
  {Boeij}}]{VK_good7}%
  \BibitemOpen
  \bibfield  {author} {\bibinfo {author} {\bibfnamefont {J.}~\bibnamefont
  {Berger}}, \bibinfo {author} {\bibfnamefont {P.}~\bibnamefont {Romaniello}},
  \bibinfo {author} {\bibfnamefont {R.}~\bibnamefont {van Leeuwen}},\ and\
  \bibinfo {author} {\bibfnamefont {P.}~\bibnamefont {Boeij}},\ }\href
  {https://doi.org/10.1103/PhysRevB.74.245117} {\bibfield  {journal} {\bibinfo
  {journal} {Phys. Rev. B}\ }\textbf {\bibinfo {volume} {74}},\ \bibinfo
  {pages} {245117} (\bibinfo {year} {2006})}\BibitemShut {NoStop}%
\bibitem [{\citenamefont {D'Agosta}\ and\ \citenamefont
  {Vignale}(2006)}]{VK_good8}%
  \BibitemOpen
  \bibfield  {author} {\bibinfo {author} {\bibfnamefont {R.}~\bibnamefont
  {D'Agosta}}\ and\ \bibinfo {author} {\bibfnamefont {G.}~\bibnamefont
  {Vignale}},\ }\href {https://doi.org/10.1103/PhysRevLett.96.016405}
  {\bibfield  {journal} {\bibinfo  {journal} {Phys. Rev. Lett.}\ }\textbf
  {\bibinfo {volume} {96}},\ \bibinfo {pages} {016405} (\bibinfo {year}
  {2006})}\BibitemShut {NoStop}%
\bibitem [{\citenamefont {Nazarov}\ \emph {et~al.}(2007)\citenamefont
  {Nazarov}, \citenamefont {Pitarke}, \citenamefont {Takada}, \citenamefont
  {Vignale},\ and\ \citenamefont {Chang}}]{VK_good9}%
  \BibitemOpen
  \bibfield  {author} {\bibinfo {author} {\bibfnamefont {V.~U.}\ \bibnamefont
  {Nazarov}}, \bibinfo {author} {\bibfnamefont {J.~M.}\ \bibnamefont
  {Pitarke}}, \bibinfo {author} {\bibfnamefont {Y.}~\bibnamefont {Takada}},
  \bibinfo {author} {\bibfnamefont {G.}~\bibnamefont {Vignale}},\ and\ \bibinfo
  {author} {\bibfnamefont {Y.-C.}\ \bibnamefont {Chang}},\ }\href
  {https://doi.org/10.1103/PhysRevB.76.205103} {\bibfield  {journal} {\bibinfo
  {journal} {Phys. Rev. B}\ }\textbf {\bibinfo {volume} {76}},\ \bibinfo
  {pages} {205103} (\bibinfo {year} {2007})}\BibitemShut {NoStop}%
\bibitem [{\citenamefont {D'Amico}\ and\ \citenamefont
  {Ullrich}(2006)}]{VK_good10}%
  \BibitemOpen
  \bibfield  {author} {\bibinfo {author} {\bibfnamefont {I.}~\bibnamefont
  {D'Amico}}\ and\ \bibinfo {author} {\bibfnamefont {C.~A.}\ \bibnamefont
  {Ullrich}},\ }\href {https://doi.org/10.1103/PhysRevB.74.121303} {\bibfield
  {journal} {\bibinfo  {journal} {Phys. Rev. B}\ }\textbf {\bibinfo {volume}
  {74}},\ \bibinfo {pages} {121303} (\bibinfo {year} {2006})}\BibitemShut
  {NoStop}%
\bibitem [{\citenamefont {Sai}\ \emph {et~al.}(2005)\citenamefont {Sai},
  \citenamefont {Zwolak}, \citenamefont {Vignale},\ and\ \citenamefont
  {Di~Ventra}}]{VK_good11}%
  \BibitemOpen
  \bibfield  {author} {\bibinfo {author} {\bibfnamefont {N.}~\bibnamefont
  {Sai}}, \bibinfo {author} {\bibfnamefont {M.}~\bibnamefont {Zwolak}},
  \bibinfo {author} {\bibfnamefont {G.}~\bibnamefont {Vignale}},\ and\ \bibinfo
  {author} {\bibfnamefont {M.}~\bibnamefont {Di~Ventra}},\ }\href
  {https://doi.org/10.1103/PhysRevLett.94.186810} {\bibfield  {journal}
  {\bibinfo  {journal} {Phys. Rev. Lett.}\ }\textbf {\bibinfo {volume} {94}},\
  \bibinfo {pages} {186810} (\bibinfo {year} {2005})}\BibitemShut {NoStop}%
\bibitem [{\citenamefont {Berger}\ \emph {et~al.}(2007)\citenamefont {Berger},
  \citenamefont {de~Boeij},\ and\ \citenamefont {van Leeuwen}}]{VK_bad}%
  \BibitemOpen
  \bibfield  {author} {\bibinfo {author} {\bibfnamefont {J.~A.}\ \bibnamefont
  {Berger}}, \bibinfo {author} {\bibfnamefont {P.~L.}\ \bibnamefont
  {de~Boeij}},\ and\ \bibinfo {author} {\bibfnamefont {R.}~\bibnamefont {van
  Leeuwen}},\ }\href {https://doi.org/10.1103/PhysRevB.75.035116} {\bibfield
  {journal} {\bibinfo  {journal} {Phys. Rev. B}\ }\textbf {\bibinfo {volume}
  {75}},\ \bibinfo {pages} {035116} (\bibinfo {year} {2007})}\BibitemShut
  {NoStop}%
\bibitem [{\citenamefont {van Faassen}\ and\ \citenamefont
  {de~Boeij}(2004{\natexlab{a}})}]{VK_bad2}%
  \BibitemOpen
  \bibfield  {author} {\bibinfo {author} {\bibfnamefont {M.}~\bibnamefont {van
  Faassen}}\ and\ \bibinfo {author} {\bibfnamefont {P.~L.}\ \bibnamefont
  {de~Boeij}},\ }\href {https://doi.org/10.1063/1.1810137} {\bibfield
  {journal} {\bibinfo  {journal} {The Journal of Chemical Physics}\ }\textbf
  {\bibinfo {volume} {121}},\ \bibinfo {pages} {10707} (\bibinfo {year}
  {2004}{\natexlab{a}})},\ \Eprint
  {https://arxiv.org/abs/https://doi.org/10.1063/1.1810137}
  {https://doi.org/10.1063/1.1810137} \BibitemShut {NoStop}%
\bibitem [{\citenamefont {van Faassen}\ and\ \citenamefont
  {de~Boeij}(2004{\natexlab{b}})}]{VK_bad3}%
  \BibitemOpen
  \bibfield  {author} {\bibinfo {author} {\bibfnamefont {M.}~\bibnamefont {van
  Faassen}}\ and\ \bibinfo {author} {\bibfnamefont {P.~L.}\ \bibnamefont
  {de~Boeij}},\ }\href {https://doi.org/10.1063/1.1697372} {\bibfield
  {journal} {\bibinfo  {journal} {The Journal of Chemical Physics}\ }\textbf
  {\bibinfo {volume} {120}},\ \bibinfo {pages} {8353} (\bibinfo {year}
  {2004}{\natexlab{b}})},\ \Eprint
  {https://arxiv.org/abs/https://doi.org/10.1063/1.1697372}
  {https://doi.org/10.1063/1.1697372} \BibitemShut {NoStop}%
\bibitem [{\citenamefont {Ullrich}\ and\ \citenamefont
  {Burke}(2004)}]{VK_bad4}%
  \BibitemOpen
  \bibfield  {author} {\bibinfo {author} {\bibfnamefont {C.~A.}\ \bibnamefont
  {Ullrich}}\ and\ \bibinfo {author} {\bibfnamefont {K.}~\bibnamefont
  {Burke}},\ }\href {https://doi.org/10.1063/1.1756865} {\bibfield  {journal}
  {\bibinfo  {journal} {The Journal of Chemical Physics}\ }\textbf {\bibinfo
  {volume} {121}},\ \bibinfo {pages} {28} (\bibinfo {year} {2004})},\ \Eprint
  {https://arxiv.org/abs/https://aip.scitation.org/doi/pdf/10.1063/1.1756865}
  {https://aip.scitation.org/doi/pdf/10.1063/1.1756865} \BibitemShut {NoStop}%
\bibitem [{\citenamefont {Berger}\ \emph {et~al.}(2005)\citenamefont {Berger},
  \citenamefont {de~Boeij},\ and\ \citenamefont {van Leeuwen}}]{VK_bad5}%
  \BibitemOpen
  \bibfield  {author} {\bibinfo {author} {\bibfnamefont {J.~A.}\ \bibnamefont
  {Berger}}, \bibinfo {author} {\bibfnamefont {P.~L.}\ \bibnamefont
  {de~Boeij}},\ and\ \bibinfo {author} {\bibfnamefont {R.}~\bibnamefont {van
  Leeuwen}},\ }\href {https://doi.org/10.1103/PhysRevB.71.155104} {\bibfield
  {journal} {\bibinfo  {journal} {Phys. Rev. B}\ }\textbf {\bibinfo {volume}
  {71}},\ \bibinfo {pages} {155104} (\bibinfo {year} {2005})}\BibitemShut
  {NoStop}%
\bibitem [{\citenamefont {Hodgson}\ \emph {et~al.}(2013)\citenamefont
  {Hodgson}, \citenamefont {Ramsden}, \citenamefont {Chapman}, \citenamefont
  {Lillystone},\ and\ \citenamefont {Godby}}]{iDEA}%
  \BibitemOpen
  \bibfield  {author} {\bibinfo {author} {\bibfnamefont {M.~J.~P.}\
  \bibnamefont {Hodgson}}, \bibinfo {author} {\bibfnamefont {J.~D.}\
  \bibnamefont {Ramsden}}, \bibinfo {author} {\bibfnamefont {J.~B.~J.}\
  \bibnamefont {Chapman}}, \bibinfo {author} {\bibfnamefont {P.}~\bibnamefont
  {Lillystone}},\ and\ \bibinfo {author} {\bibfnamefont {R.~W.}\ \bibnamefont
  {Godby}},\ }\href {https://doi.org/10.1103/PhysRevB.88.241102} {\bibfield
  {journal} {\bibinfo  {journal} {Phys. Rev. B}\ }\textbf {\bibinfo {volume}
  {88}},\ \bibinfo {pages} {241102} (\bibinfo {year} {2013})}\BibitemShut
  {NoStop}%
\bibitem [{Note2()}]{Note2}%
  \BibitemOpen
  \bibinfo {note} {The use of spinless electrons gives access to richer
  correlation for a given number of electrons. The electrons interact via the
  appropriately softened Coulomb repulsion \cite {Thesis_Hodgson} $u(x,x') =
  (|x-x'|+1)^{-1}$. We use Hartree atomic units: $m_{e}=\hbar =e=4\pi
  \varepsilon _{0}=1$.}\BibitemShut {Stop}%
\bibitem [{Note3()}]{Note3}%
  \BibitemOpen
  \bibinfo {note} {See Supplemental Material at [URL will be inserted by
  publisher] for the parameters of the model systems, and details of the
  convergence.}\BibitemShut {Stop}%
\bibitem [{\citenamefont {Ramsden}\ and\ \citenamefont
  {Godby}(2012)}]{iDEA_RE}%
  \BibitemOpen
  \bibfield  {author} {\bibinfo {author} {\bibfnamefont {J.~D.}\ \bibnamefont
  {Ramsden}}\ and\ \bibinfo {author} {\bibfnamefont {R.~W.}\ \bibnamefont
  {Godby}},\ }\href {https://doi.org/10.1103/PhysRevLett.109.036402} {\bibfield
   {journal} {\bibinfo  {journal} {Phys. Rev. Lett.}\ }\textbf {\bibinfo
  {volume} {109}},\ \bibinfo {pages} {036402} (\bibinfo {year}
  {2012})}\BibitemShut {NoStop}%
\bibitem [{\citenamefont {Thiele}\ \emph {et~al.}(2008)\citenamefont {Thiele},
  \citenamefont {Gross},\ and\ \citenamefont {K\"ummel}}]{AE_Thiele}%
  \BibitemOpen
  \bibfield  {author} {\bibinfo {author} {\bibfnamefont {M.}~\bibnamefont
  {Thiele}}, \bibinfo {author} {\bibfnamefont {E.~K.~U.}\ \bibnamefont
  {Gross}},\ and\ \bibinfo {author} {\bibfnamefont {S.}~\bibnamefont
  {K\"ummel}},\ }\href {https://doi.org/10.1103/PhysRevLett.100.153004}
  {\bibfield  {journal} {\bibinfo  {journal} {Phys. Rev. Lett.}\ }\textbf
  {\bibinfo {volume} {100}},\ \bibinfo {pages} {153004} (\bibinfo {year}
  {2008})}\BibitemShut {NoStop}%
\bibitem [{Note4()}]{Note4}%
  \BibitemOpen
  \bibinfo {note} {Our graphs show the various adiabatic and non-adiabatic KS
  potentials, etc., evaluated on the \protect \textit {exact} time-dependent
  density, so that any errors in the potentials or densities are entirely
  attributable to errors in the functionals, not the input to the
  functionals.}\BibitemShut {Stop}%
\bibitem [{\citenamefont {Tokatly}(2005{\natexlab{a}})}]{Lagrangian}%
  \BibitemOpen
  \bibfield  {author} {\bibinfo {author} {\bibfnamefont {I.~V.}\ \bibnamefont
  {Tokatly}},\ }\href {https://doi.org/10.1103/PhysRevB.71.165104} {\bibfield
  {journal} {\bibinfo  {journal} {Phys. Rev. B}\ }\textbf {\bibinfo {volume}
  {71}},\ \bibinfo {pages} {165104} (\bibinfo {year}
  {2005}{\natexlab{a}})}\BibitemShut {NoStop}%
\bibitem [{\citenamefont {Tokatly}(2005{\natexlab{b}})}]{Lagrangian2}%
  \BibitemOpen
  \bibfield  {author} {\bibinfo {author} {\bibfnamefont {I.~V.}\ \bibnamefont
  {Tokatly}},\ }\href {https://doi.org/10.1103/PhysRevB.71.165105} {\bibfield
  {journal} {\bibinfo  {journal} {Phys. Rev. B}\ }\textbf {\bibinfo {volume}
  {71}},\ \bibinfo {pages} {165105} (\bibinfo {year}
  {2005}{\natexlab{b}})}\BibitemShut {NoStop}%
\bibitem [{\citenamefont {Tokatly}(2006)}]{Lagrangian4}%
  \BibitemOpen
  \bibfield  {author} {\bibinfo {author} {\bibfnamefont {I.}~\bibnamefont
  {Tokatly}},\ }in\ \href@noop {} {\emph {\bibinfo {booktitle} {Time-Dependent
  Density Functional Theory}}},\ \bibinfo {editor} {edited by\ \bibinfo
  {editor} {\bibfnamefont {M.~A.~L.}\ \bibnamefont {Marques}}, \bibinfo
  {editor} {\bibfnamefont {C.~A.}\ \bibnamefont {Ullrich}}, \bibinfo {editor}
  {\bibfnamefont {F.}~\bibnamefont {Nogueira}}, \bibinfo {editor}
  {\bibfnamefont {A.}~\bibnamefont {Rubio}}, \bibinfo {editor} {\bibfnamefont
  {K.}~\bibnamefont {Burke}},\ and\ \bibinfo {editor} {\bibfnamefont
  {E.}~\bibnamefont {Gross}}}\ (\bibinfo  {publisher} {Springer-Verlag},\
  \bibinfo {address} {Berlin Heidelberg},\ \bibinfo {year} {2006})\
  Chap.~\bibinfo {chapter} {8}, pp.\ \bibinfo {pages} {123--136}\BibitemShut
  {NoStop}%
\bibitem [{\citenamefont {Ullrich}(2012)}]{Lagrangian5}%
  \BibitemOpen
  \bibfield  {author} {\bibinfo {author} {\bibfnamefont {C.}~\bibnamefont
  {Ullrich}},\ }\href {https://books.google.co.uk/books?id=hCNNsC4sEtkC} {\emph
  {\bibinfo {title} {Time-Dependent Density-Functional Theory: Concepts and
  Applications}}},\ Oxford Graduate Texts\ (\bibinfo  {publisher} {OUP
  Oxford},\ \bibinfo {year} {2012})\ pp.\ \bibinfo {pages}
  {465--476}\BibitemShut {NoStop}%
\bibitem [{Note5()}]{Note5}%
  \BibitemOpen
  \bibinfo {note} {The rate of change of kinetic energy is proportional to
  $u\protect \mathaccentV {dot}05F{u}$ (as in classical mechanics), and so is
  smallest when $u$ is zero. Also, if the density is moving with velocity $u$
  it will more rapidly encounter a region in which a larger non-adiabatic
  correction is required.}\BibitemShut {Stop}%
\bibitem [{Note6()}]{Note6}%
  \BibitemOpen
  \bibinfo {note} {The stated $A$ causes the wavefunction in the original frame
  to become the wavefunction in the instantaneous rest frame multiplied by
  $\protect \qopname \relax o{exp}(-iu^{2}t/2)$.}\BibitemShut {Stop}%
\bibitem [{Note7()}]{Note7}%
  \BibitemOpen
  \bibinfo {note} {The integrated absolute error, $\DOTSI \intop \ilimits@ dx \
  |n_{1}(x,t)-n_{2}(x,t)|$, expressed as a percentage of the total number of
  electrons.}\BibitemShut {Stop}%
\bibitem [{\citenamefont {Hodgson}\ \emph {et~al.}(2014)\citenamefont
  {Hodgson}, \citenamefont {Ramsden}, \citenamefont {Durrant},\ and\
  \citenamefont {Godby}}]{iDEA_SOA}%
  \BibitemOpen
  \bibfield  {author} {\bibinfo {author} {\bibfnamefont {M.~J.~P.}\
  \bibnamefont {Hodgson}}, \bibinfo {author} {\bibfnamefont {J.~D.}\
  \bibnamefont {Ramsden}}, \bibinfo {author} {\bibfnamefont {T.~R.}\
  \bibnamefont {Durrant}},\ and\ \bibinfo {author} {\bibfnamefont {R.~W.}\
  \bibnamefont {Godby}},\ }\href {https://doi.org/10.1103/PhysRevB.90.241107}
  {\bibfield  {journal} {\bibinfo  {journal} {Phys. Rev. B}\ }\textbf {\bibinfo
  {volume} {90}},\ \bibinfo {pages} {241107} (\bibinfo {year}
  {2014})}\BibitemShut {NoStop}%
\bibitem [{\citenamefont {Hodgson}(2016)}]{Thesis_Hodgson}%
  \BibitemOpen
  \bibfield  {author} {\bibinfo {author} {\bibfnamefont {M.~J.~P.}\
  \bibnamefont {Hodgson}},\ }\emph {\bibinfo {title} {Electrons in model
  nanostructures}},\ \href@noop {} {Ph.D. thesis},\ \bibinfo  {school}
  {University of York} (\bibinfo {year} {2016})\BibitemShut {NoStop}%
\bibitem [{Note8()}]{Note8}%
  \BibitemOpen
  \bibinfo {note} {Systems 2A, 2B and 2C satisfy the theorem owing to their
  symmetry, so do not form a useful test.}\BibitemShut {Stop}%
\bibitem [{Note9()}]{Note9}%
  \BibitemOpen
  \bibinfo {note} {Apart from an unimportant time-dependent
  constant.}\BibitemShut {Stop}%
\bibitem [{\citenamefont {Maitra}\ \emph {et~al.}(2002)\citenamefont {Maitra},
  \citenamefont {Burke},\ and\ \citenamefont {Woodward}}]{memory_condition}%
  \BibitemOpen
  \bibfield  {author} {\bibinfo {author} {\bibfnamefont {N.~T.}\ \bibnamefont
  {Maitra}}, \bibinfo {author} {\bibfnamefont {K.}~\bibnamefont {Burke}},\ and\
  \bibinfo {author} {\bibfnamefont {C.}~\bibnamefont {Woodward}},\ }\href
  {https://doi.org/10.1103/PhysRevLett.89.023002} {\bibfield  {journal}
  {\bibinfo  {journal} {Phys. Rev. Lett.}\ }\textbf {\bibinfo {volume} {89}},\
  \bibinfo {pages} {023002} (\bibinfo {year} {2002})}\BibitemShut {NoStop}%
\bibitem [{Note10()}]{Note10}%
  \BibitemOpen
  \bibinfo {note} {M. T. Entwistle and R. W. Godby, Data related to ``Exact
  non-adiabatic part of the Kohn-Sham potential and its fluidic
  approximation'',
  http://dx.doi.org/10.15124/8570b943-498b-4690-8044-b2208d318ef0
  (2020).}\BibitemShut {Stop}%
\end{thebibliography}%
\end{document}


\title{Exact non-adiabatic part of the Kohn-Sham potential and its fluidic approximation (Supplemental Material)} 
\date{\today}
\author{M.\ T.\ Entwistle}
\author{R.\ W.\ Godby}
\affiliation{Department of Physics, University of York, and European Theoretical Spectroscopy Facility, Heslington, York YO10 5DD, United Kingdom}

\maketitle

This document provides details on the model systems and our calculations as detailed in [URL will be inserted by publisher]. 

\section{System 1 (Two-electron Gaussian well)}
The unperturbed and perturbed external potentials are:
\begin{align}
v_{\mathrm{ext}}(x,t=0) &= -a e^{-bx^2}, \label{gaussian_gs} \\
v_{\mathrm{ext}}(x,t>0) &= -a e^{-bx^2}-\varepsilon x, \label{gaussian_td}
\end{align}
where $a=0.75$ a.u., $b=0.1$ a.u. and $\varepsilon=0.02$ a.u. The system has a spatial length of 30 a.u. and is sampled with 151 grid points, leading to a grid spacing of $\delta x = 0.2$ a.u. Real time propagation is simulated for 8 a.u. and is sampled with 1601 time points, leading to a time-step of $\delta t = 0.005$ a.u. To determine how well the fluidic approximation satisfies the zero-force theorem, we use $\delta x = 0.06$ a.u. and $\delta t = 4 \times 10^{-4}$ a.u.

\section{System 2A (Two-electron atom)}
The unperturbed and perturbed external potentials are:
\begin{align}
v_{\mathrm{ext}}(x,t=0) &= -\frac{a}{|x|+a}, \label{atom_gs}\\
v_{\mathrm{ext}}(x,t>0) &= -\frac{a}{|x|+a}+\varepsilon\cos(bx), \label{atom_td}
\end{align}
where $a=20$ a.u., $\varepsilon=0.02$ a.u. and $b=0.75$. The system has a spatial length of 40 a.u. and is sampled with 151 grid points, leading to a grid spacing of $\delta x \approx 0.27$ a.u. Real time propagation is simulated for 5 a.u. and is sampled with 1001 time points, leading to a time-step of $\delta t = 0.005$ a.u.

\section{System 2B (Three-electron atom)}
The potentials are of the same form as Eq.~(\ref{atom_gs}) and Eq.~(\ref{atom_td}) with $a=20$ a.u., $\varepsilon=0.02$ a.u. and $b=0.75$. The system has a spatial length of 50 a.u. and is sampled with 121 grid points, leading to a grid spacing of $\delta x \approx 0.42$ a.u. Real time propagation is simulated for 5 a.u. and is sampled with 1001 time points, leading to a time-step of $\delta t = 0.005$ a.u.

\section{System 2C (Strongly disrupted three-electron atom)}
The potentials are of the same form as Eq.~(\ref{atom_gs}) and Eq.~(\ref{atom_td}) with $a=20$ a.u., $\varepsilon=0.1$ a.u. and $b=0.75$. The system has a spatial length of 50 a.u. and is sampled with 121 grid points, leading to a grid spacing of $\delta x \approx 0.42$ a.u. Real time propagation is simulated for 18 a.u. and is sampled with 9001 time points, leading to a time-step of $\delta t = 0.002$ a.u.

\section{System 3 (Two-electron tunneling system)}
\begin{align}
v_{\mathrm{ext}}(x,t=0) &= \frac{a}{e^{4(-|x+4.5|+3.5)}+1} - \frac{a}{e^{4(|x-4.5|-3.5)}+1} - a, \label{tunneling_gs}\\
v_{\mathrm{ext}}(x,t>0) &= \frac{a/2}{e^{4(-|x+4.5|+3.5)}+1} - \frac{3a/2}{e^{4(|x-4.5|-3.5)}+1} - \frac{a}{2}, \label{tunneling_td}
\end{align}
where $a=0.3$ a.u. The system has a spatial length of 20 a.u. and is sampled with 301 grid points, leading to a grid spacing of $\delta x \approx 0.07$ a.u. Real time propagation is simulated for 15 a.u. and is sampled with 3001 time points, leading to a time-step of $\delta t = 0.005$ a.u.

\section{Animations}
We have produced a movie file for each system, which animates the time-evolution of the electron density: 
\begin{itemize}
    \item {\tt gaussian.mp4} (System 1)
    \item {\tt atom2.mp4} (System 2A)
    \item {\tt atom3.mp4} (System 2B)
    \item {\tt atom3\textunderscore sd.mp4} (System 2C)
    \item {\tt tunneling.mp4} (System 3)
\end{itemize}
\bibliography{Entwistle_2019}